\documentclass[10pt,a4paper]{article}
\usepackage{axodraw,amsmath,amssymb,epsfig}

\begin{document}
\begin{center}
\vspace{.5cm}{\large\bf Bloch space structure, the qutrit wave function and atom-field entanglement in three-level systems}
\end{center}
\begin{center}
\vspace{2cm}{\bf Surajit Sen,{\footnote[1]{ssen55@yahoo.com}} Mihir
Ranjan Nath {\footnote[2] {mrnath\_95@rediffmail.com}} and Tushar
Kanti Dey {\footnote[3] {tkdey54@gmail.com}}
\\Department of Physics\\
Guru Charan College\\ Silchar 788004, India \\
\vspace{.5cm} Gautam Gangopadhyay
{\footnote[4]{gautam@bose.res.in}}\\
S N Bose National Centre for Basic Sciences\\
JD Block, Sector III
\\Salt Lake City, Kolkata 700098, India}\\
\vspace{1cm}
\end{center}
\begin{center}
\hrule width 5.7in height .5pt depth 1pt
\end{center}
\begin{abstract}
We have given a novel formulation of the exact solutions for the lambda, vee and cascade three-level systems where the Hamiltonian of each configuration is expressed in the $SU(3)$ basis. The solutions are discussed from the perspective of the Bloch equation and the atom-field entanglement scenario. For the semiclassical systems, the Bloch space structure of each configuration is studied by solving the corresponding Bloch equation and it is shown that at resonance, the eight-dimensional Bloch sphere is broken up into two distinct subspaces due to the existence of a pair of quadratic constants. Because of the different structure of the Hamiltonian in the $SU(3)$ basis, the non-linear constants are found to be distinct for different configurations. We propose a possible representation of the qutrit wave function and show its equivalence with the three-level system. Taking the bichromatic cavity modes to be in the coherent state, the amplitudes of all three quantized systems are calculated by developing an Euler angle based dressed state scheme. Finally following the Phoenix-Knight formalism, the interrelation between the atom-field entanglement and population inversion for all configurations is studied and the existence of collapses and revivals of two different types is pointed out for the equidistant cascade system in particular.
\end{abstract}
\begin{center}
\hrule width 5.7in height .5pt depth 1pt
\end{center}
\vspace{4cm}
\vfill
\pagebreak
\begin{flushleft}
\large {\bf 1. Introduction}
\end{flushleft}
\par
In the atom-field interaction scenario, the level structure of an atom leads to the prediction of a wide range of experimentally verifiable coherent phenomena. Probably the most notable among them is the observation of the collapse and revival of the Rabi oscillation [1] which unequivocally proves the granular structure of the photon. This phenomenon is indeed a prediction of the Jaynes-Cummings model - an idealized two-level system consists of an atom with two distinct quantized levels interacting with a monochromatic quantized cavity field [2,3]. An immediate extension of the two-level system is the three-level system which is generally classified into lambda, vee and cascade configurations, respectively. Such configurations are in the purview of present studies because they exhibit a rich class of coherent phenomena such as, two photon coherence [4], double resonance process [5], three-level super-radiance [6], resonance Raman scattering [7], population trapping [8], tri-level echoes [9], STIRAP [10], quantum jump [11], quantum zeno effect [12], electromagnetically induced transparency [13] etc. From these studies it is quite transparently obvious that increase of the number of levels not only can generate a large number of quantum-optical effects, but also enables us to develop a suitable control mechanism which is extremely important from the experimental point of view. Thus the three-level configuration, the simplest representative of the multi-level system, demands careful inspection from time to time in its own right.
\par
It is well known that the Hamiltonians of the lambda, vee and cascade three-level systems types can be described using the atomic basis operator, $\hat{\sigma}_{\mu\nu}\equiv |\mu><\nu|$ ($\mu, \nu=1,2,3$), where the solution is carried out with the two photon resonance and the equal detuning conditions as the supplementary conditions [14,15]. Apart from this treatment, another equivalent way of dealing with the three-level system is by the Bloch equation technique, where the eight Bloch vectors are defined on the eight dimensional Bloch sphere $S^7$ [16-18]. This method was first initiated by Eberly and Hioe who pointed out the relationship of the three-level system with the $SU(3)$ group [16]. Their investigation revealed that the quadratic Casimir of the $SU(3)$ group is manifested through the existence of some nonlinear constants, which gives rise to a non-trivial structure of the Bloch space of such a system [18,19]. Later, this result was obtained by solving the pseudo-spin equation [20] and also by the Floquet theory technique [21]. However, in the Bloch equation approach, the lambda, vee and cascade three-level systems are found to be generated by changing the position of the intermediate level $E_2$, i.e., the energy levels are arranged as, $E_2>E_3>E_1$, $E_1>E_3>E_2$ and $E_3>E_2>E_1$, as shown in Fig.1 of Ref.[18]. In consequence, irrespectively of the configuration, the interaction term for any one of these three-level systems in the atomic operator basis is given by, $H_I^A=g_1|1><2|+g_2|2><3|+h.c.$ ($A=\Lambda, V$ and $\Xi$). The pitfall of having identical structures of the Hamiltonians for different configurations is that this leads to same set of non-linear constants, which is undesirable because the three-level systems are intrinsically different from one another.
\par
Apart from that, there is another reason for studying the Bloch space structure of three-level systems. In quantum information theory parlance, the qubit is "designated" by various points on the aforesaid unit Bloch sphere [22,23]. A natural extension of the qubit is the qutrit system, which is generally expressed in the computational basis: $|->$, $|0>$ and $|+>$; it has drawn wide attention in recent years [24-26]. Although there exist several suggestions for implementing the qutrit system either by treating it as the transverse spatial modes of single photons [27], or through the polarization states of the biphoton field [28,29], significant progress has been made mainly by identifying the qutrit with the three-level system driven by bichromatic laser fields [30]. However, in spite of the significant progress, a proper definition of the qutrit wavefunction and its explicit relation with the Bloch space structure of the three-level system are not available.
\par
Recently a number of quantum-optical systems have come under intense scrutiny with a view to the experimental implementation of various quantum information protocols where the concurrence is considered the most useful dynamical parameter of evolution [31]. The primary reason for such a study is that it is capable of  deciphering the nature of the entanglement between two non-local objects which is the key resource in communicating information. For example, in the phenomenon of \textsl{entanglement sudden death}, the entanglement between two non-local two-level systems is studied by considering the time evolution of the concurrence [32]. Since the two-level system is essentially an atom-field composite system, it is equally important to understand the dynamics of the entanglement between its constituents. As regards the two-level system, the correlation between the atom and the field was first considered by Gea-Banacloche in terms of purity [33] and then more systematically by Phoenix and Knight using the notion of entropy [34,35]. The latter authors also argued that the entropy can be used as an operational parameter to quantify the atom-field entanglement which is constrained by the Araki-Lieb limit. Since the three-level configuration is essentially a pump-probe system at heart, if the entanglement between two non-local such systems is realized experimentally, then it is possible to maneuver the entanglement externally. However, the entanglement dynamics of two non-local three-level systems still remains an open issue due to the lack of availability of an appropriate definition of the concurrence of the  three-level systems [31,36]. Therefore, prior to addressing the problem of the entanglement of two such non-local systems and develop a consistent theory of coherent control mechanism for engineering it by means some driving field, it is worth studying the entanglement scenario of the constituents of a single three-level system following Phoenix and Knight [34].
\par
The primary objective of this paper has two parts. First, we shall study the structure of the Bloch space for all three-level systems while taking different Hamiltonians for different configurations. Unlike in existing treatments [16,18], it is explicitly shown that if the Hamiltonians of the three-level systems are expressed in the $SU(3)$ basis with the same energy condition, this leads to distinct non-linear constants for different systems. We also give a convenient representation of the qutrit system and discuss its properties. Secondly, after developing an Euler matrix based dressed state formalism, we discuss a systematic scheme for calculating the atomic entropy following Phoenix-Knight [34] and show how the increase of the number of levels influences the atom-field entanglement and the population inversion for all three configurations.
\par
The remainder of the paper is organized as follows. In Section 2 we reconsider the Hamiltonians of the semiclassical lambda, vee and cascade three-level system types and their quantized versions expressed in the $SU(3)$ basis. Section 3 obtains the solution of the Bloch equations for the  semiclassical configurations in order to provide various non-linear constants and gives a possible representation of the wave function of the qutrit system. In Section 4, we proceed to obtain the quantized configurations by introducing an Euler angle based dressed state formalism and compare the corresponding atomic entropy with the population inversion for various initial conditions. In Section 5, we numerically study the atomic entropy with the population inversion for various initial conditions and give a possible estimate of the collapse and revival times for the cascade system in the high field approximation. Finally we summarize the main results of the paper and discuss the outlook.

\noindent
\begin{flushleft}
\large {\bf 2. The models}
\end{flushleft}
\par
The time-dependent Hamiltonian $H(t)$ can be expanded in the basis of the Lie operators as $H(t)=\sum\limits_{i=1}^N h_i(t)L_i$, where $h_i(t)$ are the linearly independent complex valued functions and the $L_i$ are the generators which satisfy the Lie algebra $[L_i,L_j]=i\omega_{ijk}L_k$. The algebraic structure of the Hamiltonian ensures the existence of a unitary operator $U(t)=\prod_{i=1}^Nexp[ig_i(t)L_i]$, where $g_i(t)$ is the scalar function which is to be evaluated for any specific model. Thus the time-dependent wave function can be evaluated either by the unitary method, $\psi(t)=U(t)\psi(0)$ [37] or by using suitable dressed state formalism [38,39]. In this Section, we develop the Hamiltonian of the semiclassical and quantized three-level systems where the atomic levels are expressed in the $SU(3)$ basis.

\noindent
It is customary to consider the three-level system as two two-level systems in the dipole approximation and the Hamiltonian of the semiclassical lambda system is given by

$${\rm H^{\Lambda}} = \hbar(\omega_1V_3+\omega_2T_3)+{\bf \hat{d}_{13}}\cdot {\bf E_1}(\Omega_1)+{\bf \hat{d}_{23}}\cdot {\bf E_2}(\Omega_2),\eqno (1)$$

\noindent
where, $\bf{\hat{d}_{13}}=(V_{+}+V_{-})$  and $\bf{\hat{d}_{23}}=(T_{+}+T_{-})$ be the dipole operators representing the transitions $1\leftrightarrow 3$ and $2\leftrightarrow 3$, and
${\bf E_1}(\Omega_1)={\bf E}_{1+}+{\bf E}_{1-}$ and ${\bf E_2}(\Omega_2)={\bf E}_{2+}+{\bf E}_{2-}$ are the bichromatic classical electric fields with mode frequencies $\Omega_1$ and $\Omega_2$, respectively. Similarly we have


$${\rm H^{V}} = \hbar(\omega_1V_3+\omega_2U_3)+{\bf \hat{d}_{13}}\cdot {\bf E_1}(\Omega_1)+{\bf \hat{d}_{12}}\cdot {\bf E_2}(\Omega_2),\eqno (2)$$

\noindent
for the vee system, where $\bf{\hat{d}_{13}}=(V_{+}+V_{-})$, $\bf{\hat{d}_{12}}=(U_{+}+U_{-})$ be the dipole operators representing transitions $1\leftrightarrow 3$ and $1\leftrightarrow 2$, respectively. Finally for the cascade system we have

$${\rm H^{\Xi}} = \hbar(\omega_1U_3+\omega_2T_3)+{\bf \hat{d}_{12}}\cdot {\bf E_1}(\Omega_1)+{\bf \hat{d}_{23}}\cdot {\bf E_2}(\Omega_2),\eqno (3)$$

\noindent
where $\bf{\hat{d}_{12}}=(U_{+}+U_{-})$ and $\bf{\hat{d}_{23}}=(T_{+}+T_{-})$ are $1\leftrightarrow 2$ and $2\leftrightarrow 3$ transitions, respectively. In Eqs.(1-3), $\hbar \omega _1 (=-{E}_1^\Lambda), \hbar \omega _2 (= -{ E}_2^\Lambda)$ and $\hbar(\omega _2 + \omega _1 )({=E}_3^\Lambda)$ are the energies of the three levels of the lambda system, $\hbar \omega _1 (={E}_1^V), \hbar \omega _2 (= { E}_2^V)$ and $-\hbar(\omega _2 + \omega _1 )({=E}_3^V)$ are those of the vee system and $\hbar \omega _1 (=-{E}_1^\Xi), \hbar (\omega _1-\omega_2) (={E}_2^\Xi)$ and $\hbar\omega _2 ({=E}_3^\Xi)$ are the energies of the cascade system, shown in Fig.1. Here we note that the energy levels of all three configurations satisfy an unique condition $E_1<E_2<E_3$, and this leads to three distinct interaction Hamiltonians, namely, $H_I^\Lambda=g_1|1><3|+g_2|3><2|+h.c.$, $H_I^V=g_1|1><2|+g_2|1><3|+h.c.$ and $H_I^\Xi=g_1|1><2|+g_2|2><3|+h.c.$, respectively. These interaction Hamiltonians in the atomic operator basis are precisely identical to the dipole interaction terms in Eqs.(1-3) which are expressed in the $SU(3)$ basis. In Eqs.(1-3), we have defined the $SU(3)$ shift vectors [40],

\begin{flushleft}
\hspace{0.5in}
 $T_{\pm}=\frac{1}{2}(\lambda_1 \pm i\lambda_2)$,
\hspace{0.35in}   $U_{\pm}=\frac{1}{2}(\lambda_{6} \pm i\lambda_{7})$,
\hspace{0.35in}   $V_{\pm}=\frac{1}{2}(\lambda_{4} \pm i \lambda_5)$,
\end{flushleft}
\begin{flushleft}
\hspace{0.5in}  $T_3=\lambda_3$,
\hspace{0.35in}  $V_3=(\sqrt{3}\lambda_8+\lambda_3)/2$
\hspace{0.5in}  $U_3=(\sqrt{3}\lambda_8-\lambda_3)/2$,\hfill (4)
\end{flushleft}

\noindent
where the Gell-Mann matrices are given by,

\begin{flushleft}
\hspace{0.5in}
 $\lambda_1   = \left[ {\begin{array}{*{20}c}
   0 & 1 & 0  \\
   1 & 0 & 0  \\
   0 & 0 & 0  \\
\end{array}} \right],
\hspace{0.35in} \lambda_ 2   = \left[ {\begin{array}{*{20}c}
   0 & -i & 0  \\
   i & 0 & 0  \\
   0 & 0 & 0  \\
\end{array}} \right],
\hspace{0.35in} \lambda_3   = \left[ {\begin{array}{*{20}c}
   1 & 0 & 0  \\
   0 & -1 & 0  \\
   0 & 0 & 0  \\
\end{array}} \right],$
\end{flushleft}

\begin{flushleft}
\hspace{0.5in}
 $\lambda_4   = \left[ {\begin{array}{*{20}c}
   0 & 0 & 1  \\
   0 & 0 & 0  \\
   1 & 0 & 0  \\
\end{array}} \right],
\hspace{0.35in} \lambda_5   = \left[ {\begin{array}{*{20}c}
   0 & 0 & -i  \\
   0& 0 & 0  \\
   i & 0 & 0  \\
\end{array}} \right],
\hspace{0.35in} \lambda_6   = \left[ {\begin{array}{*{20}c}
   0 & 0 & 0  \\
   0 & 0 & 1  \\
   0 & 1 & 0  \\
\end{array}} \right],$
\end{flushleft}

\begin{flushleft}
\hspace{0.5in}
 $\lambda_7   = \left[ {\begin{array}{*{20}c}
   0 & 0 & 0  \\
   0 & 0 & -i  \\
   0 & i & 0  \\
\end{array}} \right],
\hspace{0.25in} \lambda_8= \frac{1}{\sqrt{3}}\left[
{\begin{array}{*{20}c}
   1 & 0 & 0  \\
   0 & 1 & 0  \\
   0 & 0 & -2  \\
\end{array}} \right].\hfill (5)$
\end{flushleft}

\noindent
These matrices follow the commutation and the anti-commutation relations
\begin{flushleft} \hspace{1.5in}
$[\lambda_{i},\lambda_{j}]=2 i f_{ijk}\lambda_k$,
\hspace{0.35in}
$\{\lambda_{i},\lambda_{j}\}=\frac{4}{3}\delta_{ij}+2
d_{ijk}\lambda_k$,\hfill (6)\\
\end{flushleft}

\noindent
where $d_{ijk}$ and $f_{ijk}$ ($i,j=1,2,..,8$) represent completely symmetric and completely antisymmetric structure constants, respectively, which characterize $SU(3)$ group.

In the rotating wave approximation (RWA), the Hamiltonians of semiclassical system can be written as [39],

$$H^{\Lambda} = \hbar(\Omega
_1-\omega_1-\omega_2) V_3+\hbar(\Omega _2-\omega_1-\omega_2) {\rm
T}_3+\hbar(\Delta^{\Lambda}_1V_3+\Delta^{\Lambda}_2T_3)$$
$$+\hbar \kappa _1 V_ + \exp ( - i\Omega _1 t) +\hbar \kappa _2 T_+\exp(-i\Omega_2 t)+h.c. \quad,\eqno (7)$$

\noindent
for the lambda system,

$$H^V = \hbar(\Omega
_1-\omega_1-\omega_2) V_3+\hbar(\Omega _2-\omega_1-\omega_2) {\rm
U}_3+\hbar(\Delta^{V}_1V_3+\Delta^{V}_2U_3)$$
$$+\hbar \kappa _1 V_ + \exp ( - i\Omega _1 t) +\hbar \kappa _2 U_+\exp(-i\Omega_2 t)+h.c. \quad,\eqno (8)$$

\noindent
for the vee system, and
$$H^\Xi = \hbar(\Omega
_1-\omega_1+\omega_2) U_3+\hbar(\Omega _2+\omega_1-\omega_2) {\rm
T}_3+\hbar(\Delta^{\Xi}_1U_3+\Delta^{\Xi}_2T_3)$$
$$+\hbar \kappa _1 U_ + \exp ( - i\Omega _1 t) +\hbar \kappa _2 T_+\exp(-i\Omega_2 t)+h.c. \quad,\eqno (9)$$\\
for the cascade system, respectively. In Eq.(7-9), $\kappa_{p}$ ($p=1,2$) are the coupling parameters and $\Delta^{a}_1=(2\omega_1+\omega_2-\Omega_1)$,  $\Delta^{a}_2=(\omega_1+2\omega_2-\Omega_2)$ for $a=\Lambda, V$ and $\Delta^{\Xi}_1=(2\omega_1-\omega_2-\Omega_1)$ and $\Delta^{\Xi}_2=(2\omega_2-\omega_1-\Omega_2)$ are the respective detuning frequencies.

For the quantized system, the Hamiltonian of the lambda system is given by
$$H^\Lambda = H_0^\Lambda + H_I^\Lambda \eqno(10)$$

\noindent
where,
$$H_0^\Lambda = \hbar (\Omega _1  - \omega _1  - \omega _2 )V_3  + \hbar (\Omega _2  - \omega _1  - \omega _2 )T_3
+ \sum\limits_{j = 1}^2 {\Omega _j a_j ^\dag  } a_j\eqno(11a)$$
$$H_I^\Lambda = \hbar (\Delta _1^\Lambda  V_3  + \Delta _2^\Lambda  T_3 ) + \hbar (g_1 V_ +a_1+g_2 T_ +  a_2)+h.c.,\eqno(11b)$$

\noindent
with $a_j^\dag$ and $a_j$ ($j=1,2$) the creation and annihilation operators of two cavity modes, $g_j$ are the coupling constants and $\Omega_j$ are the mode frequencies.

Similarly the Hamiltonian of the vee system is given by
$$H^V = H_0^V + H_I^V\eqno(12)$$

\noindent
where,
$$H_0^V = \hbar (\Omega _1  - \omega _1  - \omega _2 )V_3  + \hbar (\Omega _2  - \omega _1  - \omega _2 )U_3
+ \sum\limits_{j = 1}^2 {\Omega _j a_j ^\dag  } a_j\eqno(13a)$$
$$H_I^V = \hbar (\Delta _1^V  V_3  + \Delta _2^V  U_3 ) + \hbar (g_1 V_ +a_1+g_2 U_ +  a_2)+h.c. .\eqno(13b)$$

\noindent
Finally the Hamiltonian of cascade system in the same approximation is given by

$$H^\Xi = H_0^\Xi + H_I^\Xi\eqno(14)$$

\noindent
where,
$$H_0^\Xi = \hbar (\Omega _1  + \omega _2  - \omega _1)U_3  + \hbar (\Omega _2 + \omega _1  - \omega _2 )T_3
+ \sum\limits_{j = 1}^2 {\Omega _j a_j ^\dag  } a_j\eqno(15a)$$

$$H_I^\Xi = \hbar (\Delta_1^\Xi  U_3  + \Delta _2^\Xi  T_3 ) + \hbar (g_1 U_ +a_1+g_2 T_ +  a_2)+h.c. .\eqno(15b)$$


\noindent
Using the algebra of Gell-Mann matrices it is easy to see that for the lambda and vee systems ($a=\Lambda$ and $V$)

$$[H_0^a, H_I^a ] = 0,\eqno(16)$$

\noindent
provided the two-photon resonance condition, namely,

$$\Delta_1^a=-\Delta_2^a,\eqno(17)$$

\noindent
is satisfied. Similarly for the cascade system we must have

$$[H_0^\Xi, H_I^\Xi] = 0,\eqno(18)$$

\noindent
which requires the equal detuning condition, i.e.,

$$\Delta_1^\Xi=\Delta_2^\Xi.\eqno(19)$$

\noindent
Thus the two-photon condition and the equal detuning condition ensure the commutativity of the free and interaction parts of the Hamiltonian showing that the quantized models are exactly solvable. We now proceed to discuss the Bloch space structure of all three-level configurations by exploring the possible non-linear constants obtained from the semiclassical Hamiltonians in Eqs.(7-9) written in the $SU(3)$ basis with a unique energy condition, $E_1<E_2<E_3$.

\begin{flushleft}
\large {\bf 3. Bloch equation and non-linear constants}\\
\end{flushleft}
\par
Let the solution of the Schr\"{o}dinger equation of the semiclassical three-level system described by the Hamiltonians in Eq.(7-9) be given by,
$$ \Psi^{A} (t) = C_-^A(t) \left| - \right\rangle  + C_0^A(t)
\left| 0 \right\rangle  + C_+^A(t) \left| + \right\rangle,\eqno (20)$$
where $C_-^A, C_0^A$ and  $C_+^A$ ($A=\Lambda, V$ and $\Xi$) are the amplitudes of the atomic states which satisfy the normalization condition $|C_-^A|^2+|C_0^A|^2+|C_+^A|^2=1$. In Eq.(20) the basis states of the lower, middle and upper atomic states are given by,
$$
|->=\left( \begin{array}{l}
 0 \\
 0 \\
 1 \\
 \end{array} \right), \quad |0>=\left( \begin{array}{l}
 0 \\
 1 \\
 0 \\
 \end{array} \right), \quad |+>=\left( \begin{array}{l}
 1 \\
 0 \\
 0 \\
 \end{array} \right).\\\eqno(21)
$$
The exact evaluation of the probability amplitudes enables us to calculate the density matrix defined as
$$\rho^{A}(t)=|\Psi^A(t)>\otimes<\Psi^A(t)|.\eqno (22)$$
\par
To start with, we consider the dressed wave function of the lambda system obtained by a unitary transformation
$$\tilde {\Psi}^{\Lambda}(t)=U_{\Lambda}^\dag(t) {\Psi}^{\Lambda}(0),\eqno (23)$$
where the unitary operator is given by
$$U_{\Lambda}(t) = exp[-\frac{i}{3} ((2\Omega _2 -
\Omega _1)T_3+(2 \Omega _1 - \Omega _2 )V_3 )t].\eqno (24)$$
The time-independent Hamiltonian of the lambda system described by Eq.(7) is found to be
\begin{flushleft}\hspace{0.95in}
$\tilde {H^\Lambda}(0) =
 - \hbar U_{\Lambda}^ \dagger  \dot U_{\Lambda} + U_{\Lambda}^\dagger  {H^\Lambda(t)}U_{\Lambda}$
\end{flushleft}
\begin{flushleft}
\hspace{1.4in}$ = \left[ {\begin{array}{*{20}c}
   {\frac{1}{3}\hbar( \Delta_1^\Lambda+\Delta_2^\Lambda) } & {\hbar\kappa_2  } & {\hbar\kappa_1  }  \\
   {\hbar\kappa_2  } & {\frac{1}{3}\hbar (\Delta_1^\Lambda-2\Delta_2^\Lambda) } & 0  \\
   {\hbar\kappa_1  } & 0 & {  \frac{1}{3}\hbar (\Delta_2^\Lambda-2\Delta_1^\Lambda) }   \\
\end{array}} \right].$\hfill(25)\\
\end{flushleft}
\noindent
Similarly, the unitary operator of the vee system described by the Hamiltonian in Eq.(8) is,
\begin{flushleft}
\hspace{1.5in}$ U_{V}(t) = exp[-\frac{i}{3} ((2\Omega _2 - \Omega
_1)U_3  + (2 \Omega _1 - \Omega _2 )V_3 )t],$ \hfill(26)\\
\end{flushleft}
and the corresponding time-independent Hamiltonian is given by
\begin{flushleft}
\hspace{1.0in}$ \tilde {H^V}(0) = \left[ {\begin{array}{*{20}c}
   {\frac{1}{3}\hbar (2\Delta_1^V-\Delta_2^V) } & 0 & {\hbar\kappa_1  }  \\
   0 & {\frac{1}{3}\hbar (2\Delta_2^V-\Delta_1^V) } & {\hbar\kappa_2  }  \\
   {\hbar\kappa_1  } & {\hbar\kappa_2  } & { - \frac{1}{3}\hbar (\Delta_1^V+\Delta_2^V) }   \\
\end{array}} \right].$\hfill(27)\\
\end{flushleft}
Also the Hamiltonian of the cascade system in Eq.(9) can be made time-independent by using the transformation operator
$$ U_{\Xi}(t) = exp [-\frac{i}{3} ((\Omega _1 + 2\Omega_2)T_3  + (2 \Omega _1 + \Omega _2 )U_3 )t].\eqno (28)$$\\
and the corresponding Hamiltonian is
\begin{flushleft} \hspace{1.0in}$
\tilde {H^\Xi}(0)=\left[ {\begin{array}{*{20}c}
   {\frac{1}{3}\hbar (\Delta_1^\Xi+2\Delta_2^\Xi) } & {\hbar\kappa_2  } & 0  \\
   {\hbar\kappa_2  } & {\frac{1}{3}\hbar (\Delta_1^\Xi-\Delta_2^\Xi) } & {\hbar\kappa_1  }  \\
   0 & {\hbar\kappa_1  } & { - \frac{1}{3}\hbar (2\Delta_1^\Xi+\Delta_2^\Xi) }   \\
\end{array}} \right].$\hfill(29)\\
\end{flushleft}
Thus we have three distinct Hamiltonians for three different configurations.
\par
To obtain the Bloch equation, we define the generic $SU(3)$ Bloch vectors
$$S_{i}^A(t)=Tr[\rho^A(t)\lambda_i], \eqno (30)$$
where $\rho^A$ be the density matrix given by Eq.(22) which satisfies the Liouville equation,
$$\frac{d\rho^A}{dt}=\frac{i}{\hbar}[\rho^A, \tilde{H}^A(0)].\eqno (31)$$
Eq.(30) can be equivalently expressed in terms of a Bloch vector,
$$\rho^A(t)=\frac{1}{3}({\mathbf 1}+\frac{3}{2}\sum\limits_{i = 1}^8 {S_i^A(t) \lambda_i } ).\eqno (32)$$
Substituting Eq.(32) in Eq.(31) and making use of the Hamiltonian of the requisite model we obtain the Bloch equation
$$\frac{dS_i^A}{dt}=M_{ij}^AS_j^A,\eqno (33)$$
where $M_{ij}^A$ be the eight-dimensional anti-asymmetric matrix. For the Hamiltonian of the lambda system given by Eq.(25), the matrix $M_{ij}^\Lambda$ reads
$$ M_{ij}^{\Lambda} =
\left[ {\begin{array}{*{20}c}
   0 & \Delta_2^{\Lambda} & 0 & 0  & 0 & 0 & {-\kappa_1} & 0  \\
   -{\Delta_2^{\Lambda} } & 0 & 2\kappa_2 & 0 & 0 & {-\kappa_1 } & 0 & 0  \\
   0 & {-2\kappa_2 } & 0 & 0 & {-\kappa_1 } & 0 & 0 & 0  \\
   0 & 0 & 0 & 0 & \Delta_1^{\Lambda} & 0 & {\kappa_2 } & 0  \\
   0 & 0 & {\kappa_1 } & -\Delta_1^{\Lambda} & 0 & {-\kappa_2 } &0 & {\sqrt 3 \kappa_1 }  \\
   0 & {\kappa_1} & 0 & 0 & {\kappa_2 } & 0 & (\Delta_1^{\Lambda}-\Delta_2^{\Lambda}) & 0  \\
   {\kappa_1} & 0 & 0 & {-\kappa_2} & 0 & -(\Delta_1^{\Lambda}-\Delta_2^{\Lambda}) & 0 & 0  \\
   0 & 0 & 0 & 0 & {-\sqrt 3\kappa_1 } & 0 & 0 & 0  \\
\end{array}} \right].\eqno (34)$$\\
Similarly, from the Hamiltonian in Eq.(27) we find the Bloch matrix
$$ M_{ij}^{V} =
\left[ {\begin{array}{*{20}c}
   0 & (\Delta_1^{V}-\Delta_2^{V}) & 0 & 0  & {-\kappa_2} & 0 & {-\kappa_1} & 0  \\
   -(\Delta_1^{V}-\Delta_2^{V}) & 0 & 0 & {\kappa_2} & 0 & {-\kappa_1 } & 0 & 0  \\
   0 & 0 & 0 & 0 & {-\kappa_1 } & 0 & {\kappa_2} & 0  \\
   0 & {-\kappa_2} & 0 & 0 & \Delta_1^{V} & 0 & 0 & 0  \\
   {\kappa_2} & 0 & {\kappa_1 } & -\Delta_1^{V} & 0 & 0 & 0 & {\sqrt 3 \kappa_1 }  \\
   0 & {\kappa_1 } & 0 & 0 & 0 & 0 & \Delta_2^{V} & 0  \\
   {\kappa_1} & 0 & {-\kappa_2} & 0 & 0 & -\Delta_2^{V} & 0 & {\sqrt 3 \kappa_2}  \\
   0 & 0 & 0 & 0 & {-\sqrt 3 \kappa_1 } & 0 & {-\sqrt 3 \kappa_2} & 0  \\
\end{array}} \right],\eqno (35)$$
for the vee system, and from Eq.(29) we get
$$ M_{ij}^{\Xi}  =
\left[ {\begin{array}{*{20}c}
   0 & \Delta_2^{\Xi} & 0 & 0  & -\kappa_1 & 0 & 0 & 0  \\
   -\Delta_2^{\Xi} & 0 & 2\kappa_2 & \kappa_1 & 0 & 0 & 0 & 0  \\
   0 & -2 \kappa_2 & 0 & 0 & 0 & 0 & \kappa_1 & 0  \\
   0 & -\kappa_1 & 0 & 0 & (\Delta_1^{\Xi}+\Delta_2^{\Xi}) & 0 & {\kappa_2 } & 0  \\
   \kappa_1 & 0 & 0 & -(\Delta_1^{\Xi}+\Delta_2^{\Xi}) & 0 & -\kappa_2 & 0 & 0  \\
   0 & 0 & 0 & 0 & \kappa_2 & 0 & \Delta_1^{\Xi} & 0  \\
   0 & 0 & -\kappa_1 & -\kappa_2 & 0 & -\Delta_1^{\Xi} & 0 & \sqrt{3}\kappa_1  \\
   0 & 0 & 0 & 0 & 0 & 0 & -\sqrt{3}\kappa_1 & 0  \\
\end{array}} \right],\eqno (36)$$\\
for the cascade system.
\par
The algebraic structure of $SU(3)$ group allows the existence of a set of quadratic Casimirs which will appear in the form of quadratic constants. To construct them from eight Bloch vectors, we have to search through a total of $\frac{8!}{(8-n)!n!}$ combinations which will form a tuple. However,  because of the large number of such combinations, finding the exact number of such tuple by solving the Bloch equations (34-36) is quite onerous. Therefore, to determine all possible non-linear
quadratic constants, we have developed a \verb Mathematica  program to carry out an extensive search and obtain the following results for $n=3$ and $n=5$ only.
\par
At resonance ($\Delta_1^\Lambda=0=\Delta_2^\Lambda$), for the lambda system the Bloch space is constituted of two parts, one being of two-sphere ${\mathcal S}^2$,
$$ {S_1^{\Lambda}}^2(t)+{S_4^{\Lambda}}^2(t)+{S_7^{\Lambda}}^2(t)=
{S_1^{\Lambda}}^2 (0)+{S_4^{\Lambda}}^2 (0)+{S_7^{\Lambda}}^2 (0),\eqno (37a)$$
and the other the four-sphere ${\mathcal S}^4$,
$${S_2^{\Lambda}}^2(t)+{S_3^{\Lambda}}^2(t)+{S_5^{\Lambda}}^2(t)+{S_6^{\Lambda}}^2(t)+{S_8^{\Lambda}}^2(t)=$$
$${S_2^{\Lambda}}^2 (0)+{S_3^{\Lambda}}^2 (0)+{S_5^{\Lambda}}^2 (0)+{S_6^{\Lambda}}^2 (0)+{S_8^{\Lambda}}^2 (0),\eqno (37b)$$
respectively, where ${S_i^{\Lambda}}(0)$ are the Bloch vectors at $t=0$ which are to be evaluated in terms of the probability amplitudes. On noting the fact that the density matrix can be written as $\rho^{\Lambda}(t)=U^{\dag}(t)\rho^{\Lambda}(0)U(t)$, the Bloch vector in Eq.(30) becomes,
$$S_{i}^{\Lambda}(0)=Tr[\rho^{\Lambda}(0)\lambda_i]. \eqno (38)$$
Inserting Eq.(5) and the density matrix from Eq.(22) into Eq.(38) at $t=0$, all constants $S_i^\Lambda(0)$ in Eq.(37) can be expressed in terms of probability amplitudes,
$${S_1^{\Lambda}}^2+{S_4^{\Lambda}}^2+{S_7^{\Lambda}}^2=4{C_-^{\Lambda}}(0)^2{C_0^{\Lambda}}(0)^2+4{C_0^{\Lambda}}(0)^2{C_+^{\Lambda}}(0)^2; \eqno (39a)$$
and
$$ {S_2^{\Lambda}}^2+{S_3^{\Lambda}}^2+{S_5^{\Lambda}}^2+{S_6^{\Lambda}}^2+{S_8^{\Lambda}}^2=
\frac{4}{3}(C_-^\Lambda(0)^2+C_0^\Lambda(0)^2+C_+^\Lambda(0)^2)^2$$
$$-3{C_-^{\Lambda}(0)}^2{C_0^{\Lambda}(0)}^2-3{C_0^{\Lambda}(0)}^2{C_+^{\Lambda}(0)}^2, \eqno (39b)$$
respectively. Similarly at resonance ($\Delta_1^V=0=\Delta_2^V$), for the vee system we have two Bloch spheres,
$${S_1^{V}}^2+{S_4^{V}}^2+{S_6^{V}}^2=4{C_0^{V}}(0)^2{C_+^{V}}(0)^2+4{C_+^{V}}(0)^2{C_-^{V}}(0)^2 \eqno (40a)$$
and
$${S_2^{V}}^2+{S_3^{V}}^2+{S_5^{V}}^2+{S_7^{V}}^2+{S_8^{V}}^2=
\frac{4}{3}({C_-^{V}}(0)^2+{C_0^{V}}(0)^2+{C_+^{V}}(0)^2)^2$$
$$-3{C_0^{\Lambda}(0)}^2{C_+}^{\Lambda}(0)^2-3{C_+^{\Lambda}(0)}^2{C_-^{\Lambda}(0)}^2. \eqno (40b)$$
Also at resonance ($\Delta^\Xi_1=0=\Delta^\Xi_2$), for the cascade system we have
$${S_1^{\Xi}}^2+{S_5^{\Xi}}^2+{S_6^{\Xi}}^2=4{C_-^{\Xi}}(0)^2{C_0^{\Xi}}(0)^2+4{C_0^{\Xi}}(0)^2{C_+^{\Xi}}(0)^2 \eqno (41a)$$
and
$${S_2^{\Xi}}^2+{S_3^{\Xi}}^2+{S_4^{\Xi}}^2+{S_7^{\Xi}}^2+{S_8^{\Xi}}^2=\frac{4}{3}({C_-^{\Xi}}(0)^2+{C_0^{\Xi}}(0)^2+{C_+^{\Xi}}(0)^2)^2 $$
$$-3{C_-^{\Xi}(0)}^2{C_0^{\Xi}(0)}^2-3{C_0^{\Xi}(0)}^2{C_+^{\Xi}(0)}^2,  \eqno (41b)$$
 From Eqs.(39-41) it is evident that at resonance ($\Delta_1^A=0=\Delta_2^A$), the seven-sphere ${\mathcal S}^7$ is broken into two-subspaces, namely, two-sphere ${\mathcal S}^2$ and four-sphere ${\mathcal S}^4$, repectively, each with distinct norm. We also note that for 3- and 5-tuple of the Bloch vectors constituting such two quadratic constants are in contrast to the case for [16-20], different for different three-level systems. Finally, we note that for the off-resonance case ($\Delta_1^A\neq0\neq\Delta_2^A$), the solutions of the Bloch equation of all three configurations satisfy,
$${S_1^{A}}^2+{S_2^{A}}^2+{S_3^{A}}^2+{S_4^{A}}^2+{S_5^{A}}^2+{S_6^{A}}^2+
{S_7^{A}}^2+{S_8^{A}}^2=
\frac{4}{3}({C_-^{A}}(0)^2+{C_0^{A}}(0)^2+{C_+^{A}}(0)^2)^2.\eqno (42)$$
The normalization condition, $C_-^{A}(0)^2+C_0^{A}(0)^2+C_+^{A}(0)^2=1$, readily gives that the radius-squared of the eight-dimensional Bloch sphere is $\frac{4}{3}$, regardless of the configuration.
\begin{flushleft}
{\bf 3.1. Qutrit wave function}
\end{flushleft}
\par
Having knowledge of the Bloch space of the three-level system, we now consider the qutrit wave function which is the qubit counterpart of the two-level system. we define, in the three dimensional Hilbert space $\mathcal{H}^3$, the four-parameter normalized qutrit wavefunction,
$$|q_T>=\cos\frac{\theta_0}{2}|->+\sin\frac{\theta_0}{2} \sin\frac{\theta_1}{2} \sin\frac{\theta_2}{2} e^{i\phi}|0>+$$
$$(\sin\frac{\theta_0}{2} \cos\frac{\theta_1}{2}+ i \sin\frac{\theta_0}{2} \sin\frac{\theta_1}{2} \cos\frac{\theta _2}{2})|+>,\eqno(43)$$ \\
where $\theta_0$, $\theta_1$ and $\theta_2$ lie between $0$ to $\pi$ with an arbitrary value of the phase angle $\phi$. From Eq.(43) it is evident that, the pure qutrit states $|->$, $|0>$ and $|+>$ can be obtained by taking the coordinates to be $(0,\theta_1,\theta_2)$, $(\pi,\pi,\pi)$ and $(\pi,0,\theta_2)$, respectively. This indicates that for a qutrit system, the state $|->$ may be any point on the $(\theta_1,\theta_2)$ plane, $|0>$ corresponds to a well-defined point and $|+>$ lies on a line for any value of $\theta_2$. Furthermore, it is worth noting that by taking the angles $(\theta_1,\theta_2)$ to be $(\pi,\pi)$, the qutrit wave function in Eq.(43) is reduced to a qubit wavefunction, which indicates the consistency of the qutrit wave function. Using Eq.(22), various elements of the density matrix of the qutrit wave function are given by
$$\rho_T^{11}=\cos^2\frac{\theta_1}{2},$$
$$\rho_T^{22}=\sin^2\frac{\theta_0}{2}\sin^2\frac{\theta_1}{2}\sin^2\frac{\theta_2}{2},$$
$$\rho_T^{33}=\frac{1}{4}(3+\cos\theta_1+\cos\theta_2-\cos\theta_1\cos\theta_2)\sin^2\frac{\theta_0}{2},$$
$$\rho_T^{12}={\rho_T^{21}}^{*}=\frac{1}{2}e^{i\phi}\sin\theta_0\sin\frac{\theta_1}{2}\sin\frac{\theta_2}{2},$$
$$\rho_T^{23}={\rho_T^{32}}^{*}=e^{-i\phi}\sin^2\frac{\theta_0}{2}\sin\frac{\theta_1}{2}(\cos\frac{\theta_1}{2}+i\sin\frac{\theta_1}{2}\cos\frac{\theta_2}{2})\sin\frac{\theta_2}{2},$$
$$\rho_T^{13}={\rho_T^{31}}^{*}=\frac{1}{2}\sin\theta_0(\cos\frac{\theta_1}{2} + i \sin\frac{\theta_1}{2}\cos\frac{\theta_2}{2}).\eqno (44)$$
 It is worth noting that this wavefunction is normalized, i.e., $Tr[\rho_T]=1$, which also satisfies the pure state condition $\rho_T^2=\rho_T$. Finally plucking $\rho_T$ back in Eq.(30) we obtain,
$${S_1^{A}}^2+{S_2^{A}}^2+{S_3^{A}}^2+{S_4^{A}}^2+{S_5^{A}}^2+{S_6^{A}}^2+
{S_7^{A}}^2+{S_8^{A}}^2=\frac{4}{3}, \eqno (45)$$\\
which is precisely same as Eq.(42) obtained by solving Bloch equation for different configurations of three-level systems. This establishes the complete equivalence between the qutrit with the three-level system.
\begin{flushleft}
\large {\bf 4. The Euler angle approach to dressed state formalism}
\end{flushleft}
\par
At $t=0$, the atom-field wave function of the quantized three-level system is factorized as
$${\psi_{AF}}(0) = (C_-^A(0)|-> + C_0^A(0)|0> + C_+^A(0)|+>) \otimes \sum\limits_{n, m=0}^\infty C_{n}C_{m}{|n,m>},\eqno(46)$$
\noindent
where $C_i^A$ be the atomic amplitude with level indices $i=-,0,+$ and $C_m,C_n$ are the amplitudes of the bichromatic fields which are in the coherent state with field modes $m$ and $n$, respectively. With time, the atom and field will be entangled with each other and the instantaneous atom-field wavefunction for $t>0$ is given by
$$\psi_{AF}(t)=\sum\limits_{n,m=0}^\infty{\{C_{m,n}^{-A}(t)|m,n,-> + C_{m,n}^{0A}(t)|m,n,0> + C_{m,n}^{+A}(t)|m,n,+>\}},\eqno(47)$$
where $C_{m,n}^{iA}(t)$ are the amplitude in the interaction picture which are to be evaluated.
Following the Phoenix-Knight formalism [34], the atomic entropy of the three-level system is given by
$$\mathbb{S}_{A}(t)=-Tr[\rho_{A}(t)ln\rho_{A}(t)],\eqno(48)$$
where $\rho_{A}$ be the reduced density matrix of the atomic system obtained by taking trace over the two cavity modes:
$$\rho_{A}(t)= Tr_{F_1,F_2}[\rho_{AF}(t)].\eqno(49)$$
In Eq.(49) the density matrix of the atom-field system is given by,
$$\rho_{AF}(t)=|\psi_{AF}(t)>\otimes<\psi_{AF}(t)|.\eqno(50)$$
Finally the atomic population inversion between $1 \leftrightarrow 2$, $2 \leftrightarrow 3$ and $1 \leftrightarrow 3$ levels are given by
$$W^A_{12}\equiv<T_3>=Tr_{F_1,F_2}[\rho_{AF}(t)T_3],\eqno(51a)$$
$$W^A_{23}\equiv<U_3>=Tr_{F_1,F_2}[\rho_{AF}(t)U_3],\eqno(51b)$$
$$W^A_{13}\equiv<V_3>=Tr_{F_1,F_2}[\rho_{AF}(t)V_3].\eqno(51c)$$
\noindent
In Appendix we have developed an Euler matrix based dressed state formalism to calculate the amplitudes for three possible choices of the basis states shown in Fig.1a-c. In the following subsections we present their explicit derivations for all three configurations at resonance ($\Delta_1^A=0=\Delta_2^A$).

\pagebreak

\begin{flushleft}
\large {\bf 4.1. The Lambda system}
\end{flushleft}
\par
We first consider the interaction Hamiltonian  of the lambda system given by Eq.(11) for all three possible initial conditions discussed in the Appendix. For the initial condition given by Eq.(A.1a), the Hamiltonian in the number state representation at resonance ($\Delta^\Lambda_1=0=\Delta^\Lambda_2$) is given by
$${H_{I}^{-}}^\Lambda = \left[ {\begin{array}{*{20}c}
   {0}  & {g_2\sqrt {n+1} } & g_1\sqrt {m}  \\
   {g_2\sqrt {n+1} } & 0 & {0}  \\
   g_1\sqrt {m} & {0} &  0   \\
\end{array}} \right].\eqno(52)$$\\
The eigenvalues of the Hamiltonian are $\lambda_{-1}^\Lambda = -\sqrt{g_1^2 m+g_2^2(n + 1)}(\equiv-\Omega_-^\Lambda$), $\lambda_{-2}^\Lambda=0$ and
$\lambda_{-3}^\Lambda=\sqrt{g_1^2 m+g_2^2(n + 1)} (\equiv \Omega _-^\Lambda)$ and the various elements of the orthogonal matrix in Eq.(A.8) are  \begin{flushleft}
\hspace{1.0in}$\begin{array}{l}
 {\alpha_{11}^-}^\Lambda  = \frac{1}{{\sqrt 2 }},
 \hspace{1.0cm}{\alpha_{12}^-}^\Lambda  = \frac{g_2}{\Omega _-^\Lambda} \sqrt {\frac{{n + 1}}{{2}}},
 \hspace{1.0cm}{\alpha _{13}^-}^\Lambda  = \frac{g_1}{\Omega _-^\Lambda} \sqrt {\frac{m}{{2}}},\\
 {\alpha _{21}^-}^\Lambda = 0,
 \hspace{1.3cm}{\alpha _{22}^-}^\Lambda  = \frac{g_1}{\Omega _-^\Lambda} \sqrt{m},
 \hspace{1.3cm}{\alpha _{23}^-}^\Lambda  = -\frac{g_2}{\Omega _-^\Lambda} \sqrt {n+1},\\
 {\alpha _{31}^-}^\Lambda  = -\frac{1}{{\sqrt 2 }},
 \hspace{.7cm}{\alpha _{32}^-}^\Lambda  = \frac{g_2}{\Omega _-^\Lambda} \sqrt {\frac{{n + 1}}{{2}}},
 \hspace{0.97cm}{\alpha _{33}^-}^\Lambda  = \frac{g_1}{\Omega _-^\Lambda} \sqrt {\frac{m}{{2}}}.
\end{array}$\hfill(53)\\
\end{flushleft}
The corresponding Euler angles are found to be
$$\theta_{-1}^\Lambda= \arccos[\frac{g_1}{\Omega _-^\Lambda} \sqrt {\frac{m}{{2}}}], \quad \theta_{-2}^\Lambda=
- \arccos[- g_2 \sqrt {\frac{{n + 1}}{{{\Omega^\Lambda_-}^2+
g_2^2(n + 1)}}}],$$
$$\quad \theta_{-3}^\Lambda= \arccos[- g_2\sqrt {\frac{{2n + 2}}{{{\Omega^\Lambda_-}^2+ g_2^2(n +
1)}}}].\eqno(54)$$
Thus, when the system is initially in the lower level, the corresponding time-dependent amplitudes in Eq.(A.2a) are given by
$$d_{m,n}^{-\Lambda}(t)=\frac{{g_2^2 (n+1)+g_1^2 {m} \cos\Omega_-^\Lambda t}}{{{\Omega^\Lambda_-}^2 }},\eqno(55a)$$
$$d_{m-1,n+1}^{-\Lambda}(t)=\frac{{g_1 g_2 \sqrt {m(n + 1)} (\cos\Omega _-^\Lambda t-1)}}{{{\Omega^\Lambda_-}^2 }},\eqno(55b)$$
$$d_{m-1,n}^{-\Lambda}(t)=-i \frac{{g_1 {\sqrt{m}} \sin\Omega _-^\Lambda t}}{{\Omega_-^\Lambda }}.\eqno(55c)$$
\par
For the initial condition given by Eq.(A.1b), the interaction Hamiltonian in the number state basis is
$${H_{I}^{0}}^\Lambda = \left[ {\begin{array}{*{20}c}
   {0}  & {g_2\sqrt {n} } & g_1\sqrt {m+1}  \\
   {g_2\sqrt {n} } & 0 & {0}  \\
   g_1\sqrt {m+1} & {0} &  0   \\
\end{array}} \right].\eqno(56)$$
The eigenvalues of this Hamiltonian are given by $\lambda_{01}^\Lambda = -\sqrt{g_1^2 (m+1)+g_2^2 n}(\equiv-\Omega_0^\Lambda)$, $\lambda_{02}^\Lambda=0$ and
$\lambda_{03}^\Lambda=\sqrt{g_1^2 (m+1)+g_2^2 n} (\equiv \Omega _0^\Lambda)$ and the various elements of the orthogonal matrix of Eq.(A.8) are given by
\begin{flushleft}
\hspace{1.0in}$\begin{array}{l}
 {\alpha_{11}^0}^\Lambda  =  \frac{1}{{\sqrt 2 }},
 \hspace{1.0cm}{\alpha_{12}^0}^\Lambda  = \frac{g_2}{\Omega _0^\Lambda} \sqrt {\frac{n}{{2}}},
 \hspace{1.0cm}{\alpha _{13}^0}^\Lambda  = \frac{g_1}{\Omega _0^\Lambda} \sqrt {\frac{m+1}{{2}}},\\
 {\alpha _{21}^0}^\Lambda = 0,
 \hspace{1.3cm}{\alpha _{22}^0}^\Lambda  = \frac{g_1}{\Omega _0^\Lambda} \sqrt {m+1},
 \hspace{1.1cm}{\alpha _{23}^0}^\Lambda  = -\frac{g_2}{\Omega _0^\Lambda} \sqrt {n},\\
 {\alpha _{31}^0}^\Lambda  = -\frac{1}{{\sqrt 2 }},
 \hspace{1.0cm}{\alpha _{32}^0}^\Lambda  = \frac{g_2}{\Omega _0^\Lambda} \sqrt {\frac{n}{{2}}},
 \hspace{1.0cm}{\alpha _{33}^0}^\Lambda  = \frac{g_1}{\Omega _0^\Lambda} \sqrt {\frac{m+1}{{2}}},
\end{array}$\hfill(57)\\
\end{flushleft}
which readily give the corresponding Euler's angles;
$$\theta_{01}^\Lambda= \arccos [\frac{g_1}{\Omega
_0^\Lambda} \sqrt {\frac{m+1}{{2}}}], \quad \theta_{02}^\Lambda=-
\arccos[- g_2 \sqrt {\frac{{n}}{{{\Omega^\Lambda_0}^2+ g_2^2 n
}}}], \quad \theta_{03}^\Lambda=\arccos[- g_2 \sqrt
{\frac{{2n}}{{{\Omega^\Lambda_0}^2+ g_2^2 n }}}].\eqno(58)$$
Thus, when the system is initially in the middle level, the corresponding time-dependent amplitudes in Eq.(A.2b) are given by
$$d_{m+1,n-1}^{0\Lambda}(t)=\frac{{g_1 g_2 \sqrt {(m + 1)n} (\cos\Omega_0^\Lambda t-1)}}{{{\Omega^\Lambda_0}^2 }},\eqno(59a)$$
$$d_{m,n}^{0\Lambda}(t)=\frac{{g_1^2 (m + 1) + g_2^2 n \cos\Omega_0^\Lambda t}}{{{\Omega_0^\Lambda}^2 }},\eqno(59b)$$
$$d_{m,n-1}^{0\Lambda}(t)=-i\frac{{g_2 \sqrt {n} \sin\Omega_0^\Lambda t}}{{\Omega_0^\Lambda }}.\eqno(59c)$$
Finally the interaction Hamiltonian for the initial condition Eq.(A.1c) is given by
$${H_{I}^{+}}^\Lambda = \left[ {\begin{array}{*{20}c}
   {0}  & {g_2\sqrt {n+1} } & g_1\sqrt {m+1}  \\
   {g_2\sqrt {n+1} } & 0 & {0}  \\
   g_1\sqrt {m+1} & {0} &  0   \\
\end{array}} \right],\eqno(60)$$
and the eigenvalues of the Hamiltonian are given by $\lambda_{+1}^\Lambda = -\sqrt{g_1^2 (m+1)+g_2^2 (n+1)}(\equiv -\Omega_+^\Lambda$), $\lambda_{+2}^\Lambda=0$ and $\lambda_{+3}^\Lambda=\sqrt{g_1^2 (m+1)+g_2^2 (n+1)}( \equiv \Omega_+^\Lambda)$ with various elements of the orthogonal matrix are
\begin{flushleft}
\hspace{1in}$\begin{array}{l}
 {\alpha_{11}^+}^\Lambda  =  \frac{1}{{\sqrt 2 }},
 \hspace{1.0cm}{\alpha_{12}^+}^\Lambda  = \frac{g_2}{\Omega_+^\Lambda} \sqrt {\frac{n+1}{{2}}},
 \hspace{1.0cm}{\alpha _{13}^+}^\Lambda  = \frac{g_1}{\Omega_+^\Lambda} \sqrt {\frac{m+1}{{2}}},\\
 {\alpha _{21}^+}^\Lambda = 0,
 \hspace{1.0cm}{\alpha _{22}^+}^\Lambda  = \frac{g_1}{\Omega_+^\Lambda} \sqrt{m+1},
 \hspace{1.0cm}{\alpha _{23}^+}^\Lambda  = -\frac{g_2}{\Omega_+^\Lambda} \sqrt {n+1}\\
 {\alpha _{31}^+}^\Lambda  = -\frac{1}{{\sqrt 2 }},
 \hspace{1.0cm}{\alpha _{32}^+}^\Lambda  = \frac{g_2}{\Omega_+^\Lambda} \sqrt {\frac{n+1}{{2}}},
 \hspace{0.5cm}{\alpha _{33}^+}^\Lambda  = \frac{g_1}{\Omega_+^\Lambda} \sqrt {\frac{m+1}{{2}}},
\end{array}$.\hfill(61)
\end{flushleft}
Then the corresponding Euler angles are found to be
$$\theta_{+1}^\Lambda= \arccos [\frac{g_1}{\Omega_+^\Lambda} \sqrt
{\frac{m+1}{{2}}} ], \quad \theta_{+2}^\Lambda= - \arccos[- g_2
\sqrt {\frac{{n+1}}{{{\Omega^\Lambda_+}^2+ g_2^2 (n + 1)}}}],$$
$$\theta_{+3}^\Lambda= \arccos[- g_2 \sqrt {\frac{{2n+
2}}{{{\Omega^\Lambda_+}^2+ g_2^2 (n + 1)}}}]. \eqno(62)$$

\noindent
Thus, when the system is initially in the upper level, the time-dependent amplitudes in Eq.(A.2c) are given by

$$d_{m+1,n}^{+\Lambda}(t)=-i\frac{{g_1\sqrt {m+1} \sin\Omega_+^\Lambda t}}{{\Omega_+^\Lambda}},\eqno(63a)$$

$$d_{m,n+1}^{+\Lambda}(t)=-i\frac{{g_2\sqrt {n + 1} \sin\Omega_+^\Lambda t}}{{\Omega_+^\Lambda}},\eqno(63b)$$

$$d_{m,n}^{+\Lambda}(t)=\cos\Omega_+^\Lambda t.\eqno(63c)$$

\noindent
Substituting Eqs.(55), (59) and (63) in Eq.(A.2) we obtain,

$$\psi^\Lambda_{AF}(t) = \sum\limits_{m=0,n=0}^\infty {C_m C_n \{C_-^\Lambda \frac{{g_2^2 (n+1)+g_1^2 {m} \cos\Omega^\Lambda_- t}}{{{\Omega^\Lambda_-}^2 }}}|m,n,-> + \{C_0^\Lambda {\frac{{g_1^2 (m + 1) + g_2^2 n \cos\Omega^\Lambda_0 t}}{{{\Omega^\Lambda_0}^2 }}}$$

$$+C_+^\Lambda \cos\Omega^\Lambda_+ t\}|m,n,+ >-i C_+^\Lambda {\{\frac{{g_1 \sqrt {m+1} \sin\Omega^\Lambda_+ t}}{{\Omega^\Lambda_+ }}|m+1,n,+>}+\frac{{g_2\sqrt {n+1} \sin\Omega^\Lambda_+ t}}{{\Omega^\Lambda_+ }}|m,n+1,0>\}\}$$

$$+ C_-^\Lambda \sum\limits_{m=1,n=0}^\infty { C_m C_n \{\frac{{g_1 g_2 \sqrt {m(n + 1)} (\cos\Omega^\Lambda_- t-1)}}{{{\Omega^\Lambda_-}^2 }}|m-1,n+1,0>}-i\frac{{g_1 {\sqrt{m}} \sin\Omega_-^\Lambda t}}{{\Omega_-^\Lambda}}|m - 1,n,+>\}$$

$$+C_0^\Lambda \sum\limits_{m=0,n=1}^\infty {C_m C_n \{\frac{{g_1 g_2 \sqrt {n(m + 1)} (\cos\Omega^\Lambda_0 t-1)}}{{{\Omega^\Lambda_0}^2 }}|m+1,n-1,->}- i \frac{{g_2 \sqrt {n} \sin\Omega^\Lambda_0 t}}{{\Omega^\Lambda_0 }}|m,n-1,+ >\}.\eqno(64)$$

\noindent
Finally rearranging the summation, the atom-field wavefunction for the lambda system in the entangled basis is given by

$$\psi^\Lambda_{AF}(t)=\sum\limits_{n,m=0}^\infty{\{C_{n,m}^{-\Lambda}(t)|n,m,-> + C_{n,m}^{0\Lambda}(t)|n,m,0> + C_{n,m}^{+\Lambda}(t)|n,m,+>\}},\eqno(65)$$

\noindent
where the time-dependent amplitudes are given by,

$$C_{n,m}^{-\Lambda}(t)=C_-^{\Lambda}C_{m}C_{n}\frac{g_2^2(n+1)+g_1^2 m \cos \Omega^{\Lambda}_- t}{{\Omega^{\Lambda}_-}^2}+$$
$$C_0^{\Lambda}C_{m-1}C_{n+1}\frac{{g_1 g_2 \sqrt {m(n+1)} (\cos \Omega^{\Lambda}_- t-1)}}{{\Omega^{\Lambda}_-}^2}+$$
$$-iC_+^{\Lambda}C_{m-1}C_{n}\frac{{g_1 \sqrt {m} \sin\Omega^{\Lambda}_- t}}{\Omega^{\Lambda}_-},\eqno(66a)$$

$$C_{n,m}^{0\Lambda}(t)=C_-^{\Lambda}C_{m+1}C_{n-1}\frac{g_1g_2\sqrt{n(m+1)}(\cos\Omega^{\Lambda}_0 t-1)}{{\Omega^{\Lambda}_0}^2}$$
$$+C_0^{\Lambda}C_{m}C_{n}\frac{{g_1^2(m+1)+g_2^2 n \cos\Omega^{\Lambda}_0 t}}{{{\Omega^{\Lambda}_0}^2}}$$
$$-iC_+^{\Lambda}C_{m}C_{n-1}\frac{{g_2 \sqrt {n} \sin\Omega^{\Lambda}_0 t}}{{\Omega^{\Lambda}_0}},\eqno(66b)$$

$$C_{n,m}^{+\Lambda}(t)=-i C_-^{\Lambda}C_{m+1}C_{n}\frac{g_1\sqrt{m+1}\sin{\Omega^{\Lambda}_+} t}{{{\Omega^{\Lambda}_+}}}$$
$$-i C_0^{\Lambda}C_{m}C_{n+1}\frac{{g_2\sqrt {n + 1} \sin{\Omega^{\Lambda}_+} t}}{{\Omega^{\Lambda}_+}}$$
$$+ C_+^{+\Lambda}C_{m}C_{n}cos{\Omega^{\Lambda}_+} t.\eqno(66c)$$

\pagebreak

\begin{flushleft}
\large {\bf 4.2. The Vee system}\\
\end{flushleft}
\par
For the vee system at resonance ($\Delta^V_1=0=\Delta^V_2$), the interaction Hamiltonian in Eq.(12) with the initial condition Eq.(A.1a), in the number state representation is given by
$$H_{I}^{-V} = \left[ {\begin{array}{*{20}c}
   {0}  & 0 & g_1\sqrt {m}  \\
   {0} & 0 & g_2\sqrt {n}  \\
   g_1\sqrt {m} & g_2\sqrt {n} &  0   \\
\end{array}} \right]\eqno(67).$$
The eigenvalues of the Hamiltonian are given by $\lambda_{-1}^V = -\sqrt{g_1^2 m+g_2^2 n}(\equiv -\Omega^V_-)$, $\lambda_{-2}^V=0$ and
$\lambda_{-3}^V=\sqrt{g_1^2 m+g_2^2 n} (\equiv \Omega^V_-)$ and the mixing angles are given by
$$\theta_{-1}^V = - \frac{\pi }{4}, \quad \theta_{-2}^V= \arccos [ - \frac{g_2}{\Omega^V_-} \sqrt {n}], \quad \theta_{-3}^V = - \frac{\pi }{2}.\eqno(68)$$
Thus, when the vee system is initially in the lower level, the time-dependent amplitudes are given by
$$d_{m,n}^{- V}(t) = \cos\Omega^V_- t,\eqno(69a)$$
$$d_{m,n-1}^{- V}(t) = -i\frac{{g_2 \sqrt {n} \sin\Omega^V_- t}}{{\Omega^V_-}},\eqno(69b)$$
$$d_{m-1,n}^{- V}(t)= -i \frac{{g_1 {\sqrt{m}} \sin\Omega^V_- t}}{{\Omega^V_-}}.\eqno(69c)$$
\par
For the initial condition Eq.(A.1b), the interaction Hamiltonian is given by
$${H_{I}^{0}}^V = \left[ {\begin{array}{*{20}c}
   {0}  & {0} & g_1\sqrt {m}  \\
   {0} & 0 & {g_2\sqrt {n+1}}  \\
   g_1\sqrt {m} & {g_2\sqrt {n+1} } &  0   \\
\end{array}} \right].\eqno(70)$$
The eigenvalues of the Hamiltonian are given by $\lambda_{01}^V = -\sqrt{g_1^2 m+g_2^2 (n+1)}(\equiv - \Omega^V_0)$,
$\lambda_{02}^V=0$ and $\lambda_{03}^V=\sqrt{g_1^2 m+g_2^2 (n+1)} (\equiv \Omega^V_0)$ and the corresponding Euler angles are
\hspace{1.0in}
$$\theta_{01}^V = - \frac{\pi }{4},\qquad \theta_{02}^V= \arccos [-\frac{g_2}{\Omega^V_0} \sqrt{n+1}], \qquad
\theta_{03}^V= -\frac{\pi}{2}.\eqno(71)$$
Thus, when the vee system is initially in the middle level, the time-dependent amplitudes are given by
$$d_{m,n+1}^{0 V}(t)=-i\frac{{g_2 \sqrt {n + 1} \sin\Omega^V_0 t}}{{\Omega^V_0}},\eqno(72a)$$
$$d_{m,n}^{0 V}(t)=\frac{{{g_1^2 m + g_2^2 (n+1)} \cos\Omega^V_0 t}}{{{\Omega^V_0}^2 }},\eqno(72b)$$
$$d_{m-1,n+1}^{0 V}(t)=\frac{{g_1g_2 \sqrt {m(n+1)} (-1+\cos\Omega^V_0 t)}}{{{\Omega^V_0}^2 }}.\eqno(72c)$$
\par
Finally the interaction Hamiltonian in the number state representation with the initial condition Eq.(A.1c) is given by
$${H_{I}^{+}}^V = \left[ {\begin{array}{*{20}c}
   {0}  & 0 & g_1\sqrt {m+1}  \\
   0 & 0 & {g_2\sqrt {n} }  \\
   g_1\sqrt {m+1} & {g_2\sqrt {n}} &  0   \\
\end{array}} \right]\eqno(73).$$
The eigenvalues of the Hamiltonian are given by $\lambda _{+1}^V = -\sqrt{g_1^2 (m+1) + g_2^2 n}(\equiv -\Omega^V_+)$, $\lambda_{+2}^V=0$ and
$\lambda_{+3}^V=\sqrt{g_1^2 (m+1) + g_2^2 n}\equiv(\Omega^V_+)$ and the Euler angles are given by
$$\theta_{+1}^V = - \frac{\pi }{4}, \quad \theta_{+2}^V
= \arccos [ - \frac{g_2}{\Omega^V_+} \sqrt{n}], \quad
\quad \theta_{+3}^V =- \frac{\pi }{2},\eqno(74)$$
and the corresponding time-dependent amplitudes when the system is initially in the upper level are given by
$$d_{m+1,n}^{+ V}(t)=-i\frac{{g_1\sqrt {m+1} \sin\Omega^V_+ t}}{{\Omega^V_+}},\eqno(75a)$$
$$d_{m+1,n-1}^{+ V}(t)= \frac{{g_1g_2\sqrt{m+1}\sqrt {n}(-1+\cos\Omega^V_+ t)}}{{{\Omega^V_+}^2}},\eqno(75b)$$
$$d_{m,n}^{+ V}(t)=\frac{{g_2^2 n + g_1^2(m+1)\cos\Omega^V_+ t}}{{{\Omega^V_+}^2}}.\eqno(75c)$$
Plugging Eqs.(69), (72) and (75) back in Eq.(A.12), the atom-field wavefunction in the entangled basis is given by
$$\psi^V_{AF}(t)=\sum\limits_{n,m=0}^\infty{\{C_{m,n}^{- V}(t)|n,m,-> + C_{m,n}^{0 V}(t)|n,m,0> + C_{m,n}^{+ V}(t)|n,m,+>\}},\eqno(76)$$
where ${C_i}^{V n,m}(t)$ be the required time-dependent normalized constants which are given by

$$C_{m,n}^{- V}(t)=C_-C_{m}C_{n}\cos{\Omega^V_- t}-
i C_0 C_{m}C_{n-1}\frac{{g_2 \sqrt {n} \sin \Omega^V_-
t}}{\Omega^V_-}$$
$$-iC_+ C_{m-1}C_{n}\frac{{g_1 \sqrt {m} \sin\Omega^V_- t}}{\Omega^V_-},\eqno(77a)$$

$$C_{m,n}^{0 V}(t)=-iC_-C_{m}C_{n+1}\frac{g_2 \sqrt {n+1}\sin\Omega^V_0 t}{\Omega^V_0}$$
$$+ C_0C_{m}C_{n}\frac{{g_1^2 m+g_2^2(n+1) \cos\Omega^V_- t}}{{\Omega^V_0}^2}$$
$$+ C_+C_{m-1}C_{n+1}\frac{{g_1 g_2 \sqrt {m(n+1)} (\cos\Omega^V_0 t}-1)}{{\Omega^V_0}^2},\eqno(77b)$$

$$C_{m,n}^{+ V}(t)=-iC_-C_{m+1}C_{n}\frac{g_1\sqrt{m+1}\sin\Omega_+^V t}{\Omega_+^V}
+C_0C_{m+1}C_{n-1}\frac{{g_1 g_2\sqrt{n}\sqrt{m+1}(\cos{\Omega}^V_+ t-1)}}{{{\Omega}^V_+}^2}$$

$$+ C_+C_{m}C_{n}\frac{{g_2^2 n+g_1^2 (m+1) \cos{\Omega}^V_+ t}}{{{\Omega}^V_+}^2}.\eqno(77c)$$

\begin{flushleft}
\large {\bf 4.2. The Cascade system}
\end{flushleft}
\par
For the cascade system, particularly for its equidistant configuration, we have only single quantum number with $m=n$ and $g_1=g_2\equiv g$. At resonance ($\Delta^\Xi_1=0=\Delta^\Xi_2$) with the initial condition Eq.(A.1a), the interaction Hamiltonian in Eq.(14) of the equidistant cascade system in the number state representation is given by
$${H_{I}^{-}}^\Xi = \left[ {\begin{array}{*{20}c}
   0  & g\sqrt {n-1} & 0  \\
   g\sqrt {n-1} & 0 & g\sqrt {n}  \\
   0 & g\sqrt {n} &  0   \\
\end{array}} \right].\eqno(78)$$
The eigenvalues of the Hamiltonian are given by  $\lambda _{-1}^\Xi =-g\sqrt{2n-1}(\equiv -\Omega^\Xi_-)$, $\lambda_{-2}^\Xi=0$  and
$\lambda_{-3}^\Xi=g\sqrt{2n-1} (\equiv \Omega^\Xi_-)$ and the corresponding Euler angles, which make all states entangled, are given by
$$\theta_{-1}^\Xi = -\arccos [ \sqrt{\frac{n}{4n-2}}], \quad \theta_{-2}^\Xi= -\arccos[-
\sqrt {\frac{2n-1}{3n-1}}],$$
$$\theta_{-3}^\Xi = -\arccos[-\sqrt
\frac{2n-2}{3n-2}].\eqno(79)$$
Thus, when the cascade system is initially in the lower level, the time-dependent amplitudes are given by,
$$d_{n}^{- \Xi}(t)=\frac{{{g^2 (n-1) + g^2 n} \cos\Omega^\Xi_- t}}{{{\Omega^\Xi_-}^2 }},\eqno(80a)$$
$$d_{n-1}^{- \Xi}(t)=-i\frac{{g \sqrt {n} \sin\Omega^\Xi_- t}}{{\Omega^\Xi_-}},\eqno(80b)$$
$$d_{n-2}^{- \Xi}(t)=\frac{{g^2\sqrt{(n-1)n}(-1+\cos\Omega^\Xi_- t)}}{{{\Omega^\Xi_-}^2 }}.\eqno(80c)$$
\par
Similarly for the initial condition Eq.(A.1b), the interaction Hamiltonian is given by
$${H_{I}^0}^\Xi = \left[ {\begin{array}{*{20}c}
   0  & g\sqrt {n} & 0  \\
   g\sqrt {n} & 0 & g\sqrt {n+1}  \\
   0 & g\sqrt {n+1} &  0   \\
\end{array}} \right].\eqno(81)$$
The corresponding eigenvalues of this Hamiltonian are $\lambda_{0-}^\Xi=-g\sqrt{2n+1} (\equiv -\Omega^\Xi_0)$, $\lambda_{02}^\Xi=0$
and $\lambda_{03}^\Xi=g\sqrt{2n+1} (\equiv \Omega^\Xi_0)$ and the Euler angles are found to be
$$\theta_{01}^\Xi = -\arccos [ \sqrt{\frac{n+1}{4n+2}}], \quad \theta_{02}^\Xi= -\arccos[-
\sqrt {\frac{2n+1}{3n+1}}],$$
$$\theta_{03}^\Xi = -\arccos[-\sqrt
\frac{2n}{3n+1}].\eqno(82)$$
\noindent
Thus, when the cascade system is initially in the middle level, the time-dependent amplitudes are given by
$$d_{n+1}^{0 \Xi}(t)=-i\frac{{g\sqrt {n+1} \sin\Omega^\Xi_0 t}}{{\Omega^\Xi_0}},\eqno(83a)$$
$$d_{n}^{0 \Xi}(t) = \cos\Omega^\Xi_0 t,\eqno(83b)$$
$$d_{n-1}^{0 \Xi}(t)=-i\frac{{g\sqrt {n} \sin\Omega^\Xi_0 t}}{{\Omega^\Xi_0}}.\eqno(83c)$$
\par
Finally for the initial condition Eq.(A.1c), the Hamiltonian in the number state representation is found to be
$${H_{I}^+}^\Xi = \left[ {\begin{array}{*{20}c}
   0  & g\sqrt {n+1} & 0  \\
   g\sqrt {n+1} & 0 & g\sqrt {n+2}  \\
   0 & g\sqrt {n+2} &  0   \\
\end{array}} \right].\eqno(84))$$
The eigenvalues of the Hamiltonian are given by $\lambda_{+1}^\Xi = -g\sqrt{2n+3}(\equiv -\Omega^\Xi_+)$,
$\lambda_{+2}^\Xi=0$ and $\lambda_{+3}^\Xi=g\sqrt{2n+3} (\equiv \Omega^\Xi_+)$ and the corresponding mixing angles are
$$\theta_{+1}^\Xi = -\arccos [ \sqrt{\frac{n+2}{4n+6}}], \quad \theta_{+2}^\Xi= -\arccos[-
\sqrt {\frac{2n+3}{3n+4}}],$$
$$\theta_{+3}^\Xi = -\arccos[-\sqrt
\frac{2n+2}{3n+4}],\eqno(85)$$
and the time-dependent amplitudes when the system is initially in the upper level is given by,
$$d_{n+2}^{+\Xi}(t)=\frac{{g^2 \sqrt{(n+1)(n+2)} (\cos\Omega^\Xi_+ t-1)}}{{{\Omega^\Xi_+}^2 }},\eqno(86a)$$
$$d_{n+1}^{+\Xi}(t) = -i\frac{{g \sqrt {n+1} \sin\Omega^\Xi_+ t}}{{\Omega^\Xi_+}},\eqno(86b)$$
$$d_{n}^{+\Xi}(t)=\frac{{g^2(n+2)+g^2(n+1)\cos\Omega^\Xi_+t}}{{{\Omega^\Xi_+}^2 }}.\eqno(86c)$$
Plugginging Eqs.(80), (83) and (86) back in Eq.(A.14) we obtain the atom-field wavefunction for
the cascade system in the entangled basis;

$$\psi^\Xi_{AF}(t)=\sum\limits_{n=0}^\infty{\{C_{n}^{-\Xi}(t)|n,-> + C_{n}^{0\Xi}(t)|n,0> + C_{n}^{+\Xi}(t)|n,+>\}},\eqno(87)$$
where the time-dependent normalized constants are given by,

$$C_{n}^{-\Xi}(t)=C_-^{\Xi}C_{n}^{\Xi}\frac{\{{g^2(n-1)+g^2n \cos\Omega_-^\Xi t}\}}{{{\Omega^\Xi_-}}^2}-
i C_0^{\Xi}C_{n-1}^{\Xi}\frac{g \sqrt{n} \sin\Omega^\Xi_- t}{\Omega_-^\Xi}$$
$$+C_+^{\Xi}C_{n-2}^{\Xi}\frac{{g^2 \sqrt{n(n-1)}(\cos\Omega_-^\Xi t-1)}}{{{\Omega^\Xi_-}}^2},\eqno(88a)$$

$$C_{n}^{0\Xi}(t)=-iC_-^{\Xi}C_{n+1}^{\Xi}\frac{{g\sqrt{n+1}\sin\Omega^\Xi_0 t}}{{{\Omega^\Xi_0}}}+
C_0^{\Xi}C_{n}^{\Xi}\cos\Omega_0^\Xi t$$
$$-iC_+C_{n-1}\frac{{g\sqrt{n}\sin{\Omega^\Xi_0} t}}{{{\Omega^\Xi_0}}},\eqno(88b)$$

$$C_{n}^{+\Xi}(t)=C_-^{\Xi}C_{n+2}^{\Xi}\frac{{g^2 \sqrt{(n+1)(n+2)} (\cos\Omega_+^\Xi t-1)}}{{\Omega_+^\Xi}^2}
-iC_0^{\Xi}C_{n+1}^{\Xi}\frac{{g\sqrt{n+1}\sin \Omega_+^\Xi t}}{{{\Omega_+}^\Xi}}$$
$$+C^{\Xi}_+C_{n}^{\Xi}\frac{{g^2(n+2)+g^2(n+1)\cos \Omega_+^\Xi t}}{{\Omega_+^\Xi}^2}.\eqno(88c)$$
\pagebreak
\begin{flushleft}
\large {\bf 5. Numerical results}
\end{flushleft}
\par
To understand the relationship between the atom-field entanglement and the population inversion of the quantized lambda, vee and cascade systems, we consider the following initial conditions which are relevant from the experimental point of view: a) Case I: ${C^A_-}(0)=1, {C^A_0}(0)=0, {C^A_+}(0)=0$, i.e., when the system is initially in the lower level; b) Case II: ${C^A_-}(0)=0, {C^A_0}(0)=1, {C^A_+}(0)=0$, i.e., when the system is initially in the middle level; c) Case III: ${C^A_-}(0)=0, {C^A_0}(0)=0, {C^A_+}(0)=1$, i.e., when the system is initially in the upper level. The cavity modes are taken to be in the coherent states which are characterized by the Poisson distribution, namely, $C_m=e^{-|\alpha_m|^2/2}\frac{{\alpha_m}^m}{\sqrt{m}!}$ and $C_n=e^{-|\alpha_n|^2/2}\frac{{\alpha_n}^n}{\sqrt{n}!}$, where $|\alpha_m|^2$ and $|\alpha_n|^2$ be the average photon number of two modes of the bichromatic fields. Finally, inserting the wavefunctions given by Eqs.(65), (76) and (87) in Eqs.(49-50), and then taking trace over the bi-chromatic field modes, we obtain the atomic entropy from Eq.(48) and population inversion from Eq.(51). For all configurations, the entropy lies between $0\leq\mathbb{S}_A\leq log_{2}3$ and below we consider their time evolution and compare them with the corresponding population inversion for different cases.
\par
{\bf (a) The Lambda and vee system}: Due to the existence of two modes in the Rabi frequency of the lambda and vee systems, the calculation of the collapse and revival times is non-trivial. Therefore, for these systems we give a numerical solution of the atomic entropy and population inversion only. Figs.2 and 3 give the plots of the atomic entropies and population inversions of the lambda and vee systems respectively for Cases I-III. From these graphs it is evident that during the period of collapse, the entropy reaches a minimum value which is sufficiently away from zero. This indicates that the system does not return  to the pure state during the quiescent period of collapse. Furthermore, it is noteworthy from the population inversion plots that, during collapse period, the difference in population of the involved levels is no longer zero, but maintains a constant value. Both of these features are in sharp contrast to the two-level system, where the atomic entanglement goes to almost zero during the period of collapse and two levels are equally populated, which leads to a vanishing value of the population inversion [34].
\par
{\bf (b) The Cascade system}: Figs.4-6 show the numerical plots of the entropy of the equidistant cascade system for aforementioned cases and compare them with corresponding population inversions. We note that the plot of Case-I in Fig.4 is almost identical to that for Case-III in Fig.6 showing that there exists an inversion symmetry for this system. In addition, Fig.4a, b and Fig.6a, b show that the system undergoes collapses and revivals of two distinct types with different half-width; this is not exhibited by the two level system. We now analyze the derivation of the enveloping function of the cascade system.
\par
The presence of single mode in the Rabi frequency of the cascade system enables us to get a reasonable estimate of the collapse and revival time in the high field approximation. In this approximation the mode frequency $n$ is expanded around the average photon number $\bar{n}$ which is assumed to be very large, i.e., $\bar{n} (\equiv |\alpha|^2) >>1$ [41]. Below, we have given their detail derivation for Case-I and we quote the results for other cases.
\par
{\it Case I}: The wave function of the cascade system for this case can be easily read off from Eq.(87);
$$\psi^\Xi_{AF}(t)=\psi^\Xi_{-}(t)|-> + \psi^\Xi_{0}(t)|0> + \psi^\Xi_{+}(t)|+>,\eqno(89)$$
where the field components obtained by rearranging the cavity mode are,
$$\psi^\Xi_{-}(t)=\sum\limits_{n=0}^\infty{C_{n}^{\Xi}\frac{g^2(n-1)+g^2n \cos\Omega_-^\Xi t}{{\Omega^\Xi_-}^2}}|n>,\eqno(90a)$$
$$\psi^\Xi_{0}(t)=-i\sum\limits_{n=0}^\infty{C_{n}^{\Xi}\frac{g\sqrt{n}\sin\Omega^\Xi_- t}{\Omega^\Xi_-}}|n-1>,\eqno(90b)$$
$$\psi^\Xi_{+}(t)=\sum\limits_{n=0}^\infty{C_{n}^{\Xi}\frac{g^2 \sqrt{n(n-1)} (\cos\Omega_-^\Xi t-1)}{{\Omega_-^\Xi}^2}}|n-2>.\eqno(90c)$$
Substituting Eq.(89) in Eq.(51a), the population inversion of the cascade system is,
$$W^\Xi_{12}  (t) = \sum\limits_{n=0}^\infty  C_n^2 (P_n + Q_n cos\Omega_-^\Xi + R_n cos2\Omega_-^\Xi), \eqno(91)$$
where, $P_n = \frac{(n-1)(n-2)}{2(2n - 1)^2}, Q_n = \frac{n(3n-1)}{(2n-1)^2}$ and $R_n = \frac{n^2}{2(2n-1)^2}$, respectively. To obtain the summation over the cavity modes in the aforesaid high field approximation, the Rabi frequencies in Eq.(91) can be expanded as,
$$\Omega_-^\Xi\simeq \frac{g (n-1 + \bar{n})}{\sqrt{2 \bar{n}-1}}+\mathcal{O}((n-\bar{n})^2),\eqno(92a)$$
$$\Omega_0^\Xi\simeq \frac{g (n+1+\bar{n})}{\sqrt{2 \bar{n}+1}}+\mathcal{O}((n-\bar{n})^2),\eqno(92b)$$
$$\Omega_+^\Xi\simeq \frac{g (n+3\bar{n})}{\sqrt{3 + 2 \bar{n}}}+\mathcal{O}((n-\bar{n})^2).\eqno(92c)$$
In the limit $\bar{n}>>1$, the polynomials in Eq.(91) become constant, i.e.,
$$P_n\simeq\frac{1}{8},\quad Q_n\simeq\frac{1}{2}, \quad R_n\simeq\frac{3}{8}.\eqno(93)$$
Finally, using the value of $C_n$ and inserting Eqs.(92) and (93) in Eq.(91), the summation over the cavity mode gives the population inversion between various levels for Case-I,
$$ W^\Xi_{12(-)}(t) = \frac{1}{8} + \frac{1}{2}e^{ - \bar{n} \{ 1 - \cos [\frac{{gt}}{{\sqrt {2\bar{n}  - 1} }}]\} }  \times \cos [\frac{{g(\bar{n}  - 1)t}}{{\sqrt {2\overline n  - 1} }} + \bar{n} \sin [\frac{{gt}}{{\sqrt {2\bar{n}  - 1} }}]] +  $$
$$ \frac{3}{8} e^{ - \bar{n} \{ 1 - \cos [\frac{{2gt}}{{\sqrt {2\bar{n}  - 1} }}]\} }  \times \cos [\frac{{2g(\bar{n}  - 1)t}}{{\sqrt {2\bar{n}  - 1} }} + \bar{n} \sin[\frac{{2gt}}{{\sqrt {2\bar{n}  - 1} }}]],\eqno(94a) $$

$$ W^\Xi_{23( - )}(t) = -\frac{1}{8} + \frac{1}{2}e^{ - \bar{n} \{ 1 - \cos [\frac{{gt}}{{\sqrt {2\bar{n} -1} }}]\} }  \times \cos [\frac{{g(2\bar{n} - 1)t}}{{\sqrt {\bar{n} - 1} }} + \bar{n}\sin [\frac{{gt}}{{\sqrt {2\bar{n} - 1} }}]]-$$
$$\frac{3}{8}e^{ - \bar{n}\{ 1 - \cos [\frac{{2gt}}{{\sqrt {2\bar{n} - 1}}}]\} }  \times \cos [\frac{{2g(\bar{n} - 1)t}}{{\sqrt {2\bar{n} - 1} }} + \bar{n} \sin [\frac{{2gt}}{{\sqrt {2\bar{n} - 1} }}]],\eqno(94b)$$

$$ W^\Xi_{13(-)}(t) = e^{ - \bar{n} \{ 1 - \cos [\frac{{gt}}{{\sqrt {2\bar{n}  - 1} }}]\} }  \times \cos [\frac{{g(\bar{n} - 1)t}}{{\sqrt {2\bar{n} - 1} }} + \bar{n} \sin[\frac{{gt}}{{\sqrt {2\bar{n} - 1} }}]]. \eqno(94c)$$
From the two enveloping functions given by Eqs.(94a) and (94b), we note that there are two distinct times of collapse and times of revival. In the limit $\bar{n}>>1$, the two times of collapse are
$$t^c_{1(-)} = \frac{\sqrt{2(2\bar{n}-1)}}{g\sqrt{\bar{n}}} \simeq \frac{2}{g},\eqno(95a)$$,
$$t^c_{2(-)} = \frac{t^c_{1(-)}}{2} = \frac{\sqrt{2\bar{n}-1}}{g\sqrt{2\bar{n}}} \simeq \frac{1}{g},\eqno(95b)$$
and similarly, two times of revival are
$$t^r_{1(-)} = \frac{2\sqrt{2\bar{n}-1}}{g}\pi \simeq \frac{2\sqrt{2\bar{n}}}{g}\pi,\eqno(96a)$$
$$t^r_{2(-)} = \frac{t^r_{1(-)}}{2} = \frac{\sqrt{2\bar{n}-1}}{g}\pi \simeq \frac{\sqrt{2\bar{n}}}{g}\pi.\eqno(96b)$$
In contrast, the population inversion between the levels ${1}\leftrightarrow{3}$ represented by Eq.(94c) possesses single time of collapse and revival given by $t^c_{1(-)}$ and $t^r_{1(-)}$, respectively.
\par
Proceeding in a similar way, the population inversion for Case-II are given by
$$ W^\Xi_{12(0)}(t) = - \frac{3}{8} - \frac{5}{8} e^{ - \bar{n} \{ 1 - \cos [\frac{{gt}}{{\sqrt {2\bar{n} + 1} }}]\} }  \times \cos [\frac{{g(\bar{n} + 1)t}}{{\sqrt {2\bar{n} + 1} }} + \bar{n} \sin [\frac{{gt}}{{\sqrt {2\bar{n} + 1} }}]],\eqno(97a)$$
$$ W^\Xi_{23(0)}(t) = \frac{1}{4} + \frac{3}{4} e^{ - \bar{n} \{ 1 - \cos [\frac{{gt}}{{\sqrt {2\bar{n}  + 1} }}]\} }  \times \cos [\frac{{2g(\bar{n} + 1)t}}{{\sqrt {2\bar{n}  + 1} }} + \bar{n} \sin [\frac{{2gt}}{{\sqrt {2\bar{n} + 1} }}]],\eqno(97b)$$
$$ W^\Xi_{13(0)}(t) = -\frac{1}{8} + \frac{1}{8} e^{ - \bar{n} \{ 1 - \cos [\frac{{gt}}{{\sqrt {2\bar{n} + 1} }}]\} }  \times \cos [\frac{{2g(\bar{n} + 1)t}}{{\sqrt {2\bar{n} + 1} }} + \bar{n}\sin [\frac{{2gt}}{{\sqrt {2\bar{n} + 1} }}]].\eqno(97c)$$
In this case we see that the collapse and revival time are given by

$$t^c_{1(0)} = \frac{\sqrt{2(2\bar{n}+1)}}{g\sqrt{\bar{n}}} \simeq \frac{2}{g},\eqno(98a)$$,
$$t^r_{1(0)} = \frac{2\sqrt{2\bar{n}+1}}{g}\pi \simeq \frac{2\sqrt{2\bar{n}}}{g}\pi.\eqno(98b)$$
Finally for Case-III, which is similar to Case-I, the population inversions among various levels are,

$$ W^\Xi_{12(+)}(t) = \frac{1}{8} - \frac{1}{2}e^{ - \bar{n} \{ 1 - \cos [\frac{{gt}}{{\sqrt {3\bar{n} + 2} }}]\} }  \times \cos [\frac{{g(\bar{n} + 3)t}}{{\sqrt {2\bar{n} + 3} }} + \bar{n} \sin [\frac{{gt}}{{\sqrt {2\bar{n} + 3} }}]] + $$
$$\frac{3}{8} e^{ - 2\bar{n} \{ 1 - \cos [\frac{{2gt}}{{\sqrt {\bar{n} + 3} }}]\} }  \times \cos [\frac{{2g(\bar{n} + 3)t}}{{\sqrt {2\bar{n} + 3} }} + \bar{n} \sin [\frac{{2gt}}{{\sqrt {2\bar{n} + 3} }}]], \eqno(99a)$$

$$ W^\Xi_{23(+)}(t) = -\frac{1}{8} - \frac{1}{2}e^{ - \bar{n} \{ 1 - \cos [\frac{{gt}}{{\sqrt {2\bar{n}  + 3} }}]\} }  \times \cos [\frac{{g(\bar{n} + 3)t}}{{\sqrt {2\bar{n} + 3} }} + \bar{n}\sin [\frac{{gt}}{{\sqrt {2\bar{n} + 3} }}]]-$$

$$ \frac{3}{8}e^{ - \bar{n}\{ 1 - \cos [\frac{{2gt}}{{\sqrt {2\bar{n}+3}}}]\} }  \times \cos [\frac{{2g(\bar{n} + 3)t}}{{\sqrt {2\bar{n} + 3} }} + \bar{n} \sin [\frac{{2gt}}{{\sqrt {\bar{n} - 1} }}]],\eqno(99b)$$

$$ W^\Xi_{13(+)}(t) = - e^{ - \bar{n} \{ 1 - \cos [\frac{{gt}}{{\sqrt {2\bar{n}  + 3} }}]\} }  \times \cos [\frac{{2g(\bar{n} + 3)t}}{{\sqrt {2\bar{n}  + 3} }} + \bar{n} \sin [\frac{{gt}}{{\sqrt {2\bar{n} + 3} }}]], \eqno(99c)$$
and two times of collapse are found to be
$$t^c_{1(+)} = \frac{\sqrt{2(2\bar{n}+3)}}{g\sqrt{\bar{n}}} \simeq \frac{2}{g},\eqno(100a)$$,
$$t^c_{2(+)} = \frac{t^c_{1+}}{2} = \frac{\sqrt{2\bar{n}+3}}{g\sqrt{2\bar{n}}} \simeq \frac{1}{g},\eqno(100b)$$
with the corresponding revival times,
$$t^r_{1(+)} = \frac{2\sqrt{2\bar{n}+3}}{g}\pi \simeq \frac{2\sqrt{2\bar{n}}}{g}\pi,\eqno(101a)$$
$$t^r_{2(+)} = \frac{t^r_{1+}}{2} = \frac{\sqrt{2\bar{n}+3}}{g}\pi \simeq \frac{\sqrt{2\bar{n}}}{g}\pi,\eqno(101b)$$
respectively. Thus in the high field approximation, two distinct collapse and revival situations with two different time scales are present in the numerical studies also.
\pagebreak
\begin{flushleft}
\large {\bf 6. Conclusion}
\end{flushleft}
\par
In this paper we have reconsidered the exact solution for all three semiclassical and quantized three-level systems where the Hamiltonians are written in the $SU(3)$ basis. The semiclassical systems are solved by using the Bloch equation approach and it is shown that at zero detuning, the eight dimensional Bloch sphere is broken into two disjoint sectors due to the existence of two distinct nonlinear constants. We emphasize that such pair of nonlinear constants, which are different for different configurations, were not considered in earlier treatments where the Hamiltonians no longer obeyed a unique energy condition demanded by the $SU(3)$ symmetry [16-20]. A significant outcome of our approach is that we have given a possible representation of the qutrit wavefunction which not only satisfies the pure state condition, but also has established its equivalence with the three-level system. Taking the cavity modes to be in the coherent state, an Euler matrix based dressed state formalism is developed for finding the amplitudes of the quantized three-level systems and the atom-field entanglement scenario is studied by comparing the atomic entropies with corresponding population inversions for various initial conditions. We have explicitly derived the envelope  functions of the equidistant cascade system for all cases and have shown that there are two distinct patterns of collapse and revival. The collapse and revival times of both of these patterns are calculated in the high field approximation. Although the conventional quantum information processing has made significant progress with qubit as the primitive entity, a study of the qutrit wavefunction is a necessary prerequisite for developing the quantum information theory beyond the qubit. More exploration in this direction may give some interesting results which are within reach of future cavity experiments.
\vfill
\begin{flushleft}
\large {\bf Acknowledgements}
\end{flushleft}
We thank anonymous referee for helpful suggestions. SS is thankful to Department of Science and Technology, New Delhi for partial support and thanks S N Bose National Centre for Basic Sciences, Kolkata, for supporting his visit there. TKD also thanks S N Bose National Centre for Basic Sciences, Kolkata, for supporting his visit there through the Associateship Program.
\pagebreak
\begin{flushleft}
\large {\bf Appendix}
\end{flushleft}
\par
For completeness, in this appendix we shall present the Euler matrix based dressed state formalism to calculate the time-dependent amplitudes [38,39]. To find the atom-field wavefunction in the interaction picture, let us note that there are three possibilities for defining the initial state wavefunction with the quantum numbers $(n,m)$ shown in Fig.1a,b,c for the lambda, vee and cascade configuration types, respectively. Once this quantum number is assigned to any one particular level, then the photon quantum number of other two levels will be determined by the selection rules $|\Delta{n}|=1$ and $|\Delta{m}|=1$, respectively. Thus at $t=0$ there are three possible choices for the three initial wavefunctions: (i) when the system is initially in the lower level,
$$\psi^{-A}_{AF}(0) = C_-^A|0>\otimes\sum\limits_{m=0}^\infty \sum\limits_{n=0}^\infty C_{m}C_{n}|m,n>,\eqno(A.1a)$$
(ii) when the system is initially in the middle level,
$$\psi^{0A}_{AF}(0) = C_0^A|0>\otimes\sum\limits_{m=0}^\infty \sum\limits_{n=0}^\infty C_{m}C_{n}|m,n>,\eqno(A.1b)$$
(iii) when the system is initially in the upper level,
$$\psi^{+A}_{AF}(0) = C_+^A|+>\otimes\sum\limits_{m=0}^\infty \sum\limits_{n=0}^\infty C_{m}C_{n}|m,n>.\eqno(A.1c)$$
Taking above states as the initial states, the instantaneous wave function for the lambda, vee and cascade models can be obtained. Below we shall outline the derivation of the amplitudes of the atom-field wavefunction of the lambda system in detail and the remaining two are similar.

\par{\bf (a) The lambda system:}

\noindent
In the interaction picture,the atom-field wavefunctions for the lambda system are given by,
$$\psi^{-\Lambda}_{AF}(t) = C_-^\Lambda \sum\limits_{n=0}^\infty { \{\sum\limits_{m=0}^\infty C_{m}C_{n} d_{m,n}^{-\Lambda}(t)|m,n,->+\sum\limits_{m=1}^\infty C_{m}C_{n} d_{m-1,n+1}^{-\Lambda}(t)|m-1,n+1,0>}$$
$$+\sum\limits_{m=1}^\infty C_{m}C_{n} d_{m-1,n}^{-\Lambda}(t)|m-1,n,+>\},\eqno(A.2a)$$
$$\psi^{0\Lambda}_{AF}(t) = C_0^\Lambda {\sum\limits_{m=0}^\infty \{\sum\limits_{n=1}^\infty C_{m}C_{n} d_{m+1,n-1}^{0\Lambda}(t)|m+1,n-1,0>+C_{m}C_{n} \sum\limits_{n=0}^\infty d_{m,n}^{0\Lambda}(t)|m,n,+>}$$
$$+\sum\limits_{n=1}^\infty C_{m}C_{n} d_{m,n-1}^{0\Lambda}(t)|m,n-1,+>\},\eqno(A.2b)$$
$$\psi^{+\Lambda}_{AF}(t) = C_+^\Lambda {\sum\limits_{m=0,n=0}^\infty \{C_{m}C_{n} d_{m+1,n}^{+\Lambda}(t)|m+1,n,->+C_{m}C_{n}d_{m,n+1}^{+\Lambda}(t)|m,n+1,0>}$$
$$+C_{m}C_{n}d_{m,n}^{+\Lambda}(t)|m,n,+>\},\eqno(A.2c)$$

\noindent
where $d_{m,n}^{+\Lambda}(t)$ etc represent time-dependent amplitudes which are to be evaluated for three possible choices of basis states shown in Fig.1a. Using superposition principle, the total wave function of the atom-field system is given by

\noindent
$$\psi^\Lambda_{AF}(t)=\psi^{-\Lambda}_{AF}(t)+\psi^{0\Lambda}_{AF}(t)+\psi^{+\Lambda}_{AF}(t).\eqno(A.3)$$

\noindent
Finally, rearranging the summation of the eigen modes, the atom-field wavefunction is given by

$$\psi^\Lambda_{AF}(t)=\sum\limits_{n,m=0}^\infty{\{C_{m,n}^{- \Lambda}(t)|n,m,-> + C_{m,n}^{0 \Lambda}(t)|n,m,0> + C_{m,n}^{+ \Lambda}(t)|n,m,+>\}},\eqno(A.4)$$
where $C_{m,n}^{+ \Lambda}(t)$ etc are the time-dependent amplitudes in the atom-field entangled basis given by
$$C_{m,n}^{-\Lambda}(t)=C_-^\Lambda C_{m}C_{n}d_{m,n}^{-\Lambda}(t)+C_0^\Lambda C_{m-1}C_{n+1}d_{m-1,n+1}^{0\Lambda}(t)+C_+^\Lambda C_{ m-1}C_{n}d_{m-1,n}^{+\Lambda}(t),\eqno(A.5a)$$
$$C_{m,n}^{0 \Lambda}(t)=C_-^\Lambda C_{m+1}C_{n-1}d_{m+1,n-1}^{-\Lambda}(t)+C_0 ^\Lambda C_{m}C_{n}d_{m,n}^{0\Lambda}(t)+C_+^\Lambda C_{m}C_{n-1}d_{m,n-1}^{+\Lambda}(t),\eqno(A.5b)$$
$$C_{m,n}^{+ \Lambda}(t)=C_-^\Lambda C_{m+1}C_{n}d_{m+1,n}^{-\Lambda}(t)+C_0^\Lambda C_{m}C_{n+1}d_{m,n+1}^{0\Lambda}(t)+C_+^\Lambda C_{m}C_{n}d_{m,n}^{+\Lambda}(t).\eqno(A.5c)$$

To derive above expressions, we note that the interaction Hamiltonian of the lambda system in the number state basis is given by $(H_I^{i\Lambda})_{m,n}=<\psi_{AF}^{i\Lambda}|H_I^\Lambda|\psi_{AF}^{i\Lambda}>$ and it can be diagonalized as

$$diag(\lambda^\Lambda_{i3}, \lambda^\Lambda_{i2}, \lambda^\Lambda_{i1})=T_{m,n}^i (H_I^{i\Lambda})_{m,n} {T_{m,n}^i}^{-1}\eqno(A.6)$$

\noindent
where $\lambda^\Lambda_{i1}, \lambda^\Lambda_{i2}$ and $\lambda^\Lambda_{i3}$ ($i = -, 0, +$) be the eigen values of the Hamiltonian and $T_{m,n}^i$ be the transformation matrix which gives the dressed wave function from the bare states by the orthogonal rotation,

\begin{flushleft}
\hspace{1.5in}$\left[ {\begin{array}{*{20}c}
   {|m,n,1> }  \\
   {|m,n,2> }  \\
   {|m,n,3> }  \\
\end{array}} \right] = T_{m,n}^i\left[ {\begin{array}{*{20}c}
   {|m-1,n,-> }  \\
   {|m-1,n+1,0> }  \\
   {|m,n,+> }  \\
\end{array}} \right].$\hfill(A.7)
\end{flushleft}

In Eq.(A.7), the orthogonal matrix may be parameterized by the Euler matrix

\begin{flushleft}
\hspace{1.5in} $T_{m,n}^i = \left[ {\begin{array}{*{20}c}
   {\alpha_{11}^i} & {\alpha_{12}^i} & {\alpha_{13}^i}  \\
   {\alpha_{21}^i} & {\alpha_{22}^i} & {\alpha_{23}^i}  \\
   {\alpha_{31}^i} & {\alpha_{32}^i} & {\alpha_{33}^i}  \\
\end{array}} \right],$\hfill(A.8)
\end{flushleft}

\noindent
where,

\begin{flushleft}
\hspace{1.5in}
 $\begin{array}{l}
\alpha_{11}^i  = c_{1}^i c_{2}^i-c_{3}^i s_{2}^i s_{1}^i  \\
 \alpha_{12}^i  = c_{1}^i s_{2}^i + c_{3}^i c_{2}^i s_{1}^i  \\
 \alpha_{13}^i  = s_{1}^i s_{3}^i  \\
 \alpha_{21}^i  =  - s_{1}^i c_{2}^i - c_{3}^i s_{2}^i c_{1}^i \\
 \alpha_{22}^i  =  - s_{1}^i s_{2}^i + c_{3}^i c_{2}^i c_{1}^i \\
 \alpha_{23}^i  = c_{1}^i s_{3}^i  \\
 \alpha_{31}^i  = s_{3}^i s_{2}^i  \\
 \alpha_{32}^i  =  - s_{3}^i c_{2}^i  \\
 \alpha_{33}^i  = c_{3}^i. \\
 \end{array}$\hfill(A.9)
\end{flushleft}

\noindent
with $s_p^i=\sin\theta_{ip}^{\Lambda}$ and $c_p^i=\cos\theta_{ip}^{\Lambda}$ ($p=1,2,3$) three mixing angles which are to be evaluated for the above choices of the basis states shown in Fig.1. For choice-I, the time-dependent probability amplitudes in Eq.(A.2a) obtained from the initial wavefunction Eq.(A.1a) are given by the transformation,
\begin{flushleft}
\hspace{1in} $\left[ {\begin{array}{*{20}c}
   {d_{m-1,n}^{- \Lambda} (t)}  \\
   {d_{m-1,n+1}^{-\Lambda} (t)}  \\
   {d_{m,n}^{-\Lambda}(t)}  \\
\end{array}} \right] = {T_{m,n}^-}^{-1}\left[ {\begin{array}{*{20}c}
   {e^{ - i{\lambda^\Lambda_{-3}} t} } & 0 & 0  \\
   0 & {e^{ - i\lambda^\Lambda_{-2} t} } & 0  \\
   0 & 0 & {e^{i\lambda^\Lambda_{-1} t} }  \\
\end{array}} \right]{T_{m,n}^-}\left[ {\begin{array}{*{20}c}
   {0}  \\
   {0}  \\
   {1}  \\
\end{array}} \right].$\hfill(A.10a)
\end{flushleft}
Similarly, for choice-II in Eq.(A.1b), the corresponding amplitudes in Eq.(A.2b) are obtained by
\begin{flushleft}
\hspace{1in} $\left[ {\begin{array}{*{20}c}
   {d_{m,n-1}^{0 \Lambda}(t)}  \\
   {d_{m,n}^{0 \Lambda}(t)}  \\
   {d_{m+1,n-1}^{0 \Lambda}(t)}  \\
\end{array}} \right] = {T_{m,n}^0}^{-1}\left[ {\begin{array}{*{20}c}
   {e^{ - i\lambda^\Lambda_{0 3} t} } & 0 & 0  \\
   0 & {e^{ - i\lambda^\Lambda_{0 2} t} } & 0  \\
   0 & 0 & {e^{i\lambda^\Lambda_{01} t} }  \\
\end{array}} \right]{T_{m,n}^0}\left[ {\begin{array}{*{20}c}
   {0}  \\
   {1}  \\
   {0}  \\
\end{array}} \right],$\hfill(A.10b)
\end{flushleft}

\noindent
and finally for choice-III, the amplitudes in Eq.(A.2c) are obtained from Eq.(A.1c),

\begin{flushleft}
\hspace{1in} $\left[ {\begin{array}{*{20}c}
   {d_{m,n}^{+ \Lambda} (t)}  \\
   {d_{m,n+1}^{+ \Lambda}(t)}  \\
   {d_{m+1,n}^{+ \Lambda}(t)}  \\
\end{array}} \right] = {T_{m,n}^+}^{ - 1} \left[ {\begin{array}{*{20}c}
   {e^{ - i\lambda^\Lambda_{+3} t} } & 0 & 0  \\
   0 & {e^{ - i\lambda^\Lambda_{+2} t} } & 0  \\
   0 & 0 & {e^{i\lambda^\Lambda_{+1} t} }  \\
\end{array}} \right]{T_{m,n}^+}\left[ {\begin{array}{*{20}c}
   {1}  \\
   {0}  \\
   {0}  \\
\end{array}} \right].$\hfill(A.10c)
\end{flushleft}

\noindent
Finally plugging Eqs.(A.10) back into Eq.(A.2) we obtain required time-dependent amplitudes of the lambda system given by Eq.(A.5).

{\bf (b) The vee system:}

\noindent
Proceeding in similar way, the atom-field wavefunction for the vee system is given by,
$${\psi}^V_{AF}(t)=\sum\limits_{n,m=0}^\infty{\{C_{n,m}^{-V}(t)|n,m,-> + C_{n,m}^{0V}(t)|n,m,0> + C_{n,m}^{+V}(t)|n,m,+>\}},\eqno(A.11)$$

\noindent
where the time-dependent amplitudes $C_{m,n}^{i V}(t)$ in the atom-field entangled basis are given by

$$C_{m,n}^{-V}(t)=C_-^{V}C_{m}C_{n}d_{m,n}^{-V}(t)+C_0^{V}C_{m}C_{n-1}d_{m,n-1}^{0V}(t)+C_+^{V}C_{m-1}C_{n}d_{m-1,n}^{+V}(t),\eqno(A.12a)$$
$$C_{m,n}^{0 V}(t)=C_-^{V}C_{m}C_{n+1}d_{m,n+1}^{- V}(t)+C_0^{V}C_{m}C_{n}d_{m,n}^{0 V}(t)+C_+^{V}C_{m-1}C_{n+1}d_{m-1,n+1}^{+ V}(t),\eqno(A.12b)$$
$$C_{m,n}^{+ V}(t)=C_-^{V}C_{m+1}C_{n}d_{m+1,n}^{- V}(t)+C_0^{V}C_{m+1}C_{n-1}d_{m+1,n-1}^{0 V}(t)+C_+^{V}C_{m}C_{n}d_{m,n}^{+ V}(t).\eqno(A.12c)$$

{\bf (c) The cascade system:}

Similarly, for the equidistant cascade system the wavefunction is found to be,
$${\psi}^\Xi_{AF}(t)=\sum\limits_{n=0}^\infty{\{C_{n}^{-\Xi}(t)|n,-> + {C_{n}^{0\Xi}}(t)|n,0> + C_{n}^{+\Xi}(t)|n,+>\}},\eqno(A.13)$$
where the time-dependent amplitudes are given by
$$C_{n}^{-\Xi}(t)=C_-^{\Xi}C_{n}d_{n}^{-\Xi}(t)+C_0^{\Xi}C_{n+1}d_{n+1}^{0\Xi}(t)+C_+^{\Xi}C_{n+2}d_{n+2}^{+\Xi}(t),\eqno(A.14a)$$
$$C_{n}^{0\Xi}(t)=C_-^{\Xi}C_{n-1}d_{n-1}^{-\Xi}(t)+C_0^{\Xi}C_{n}d_{n}^{0\Xi}(t)+C_+^{\Xi}C_{n+1}d_{n+1}^{+\Xi}(t),\eqno(A.14b)$$
$$C_{n}^{+\Xi}(t)=C_-^{\Xi}C_{n-2}d_{n-2}^{-\Xi}(t)+C_0^{\Xi}C_{n-1}d_{n-1}^{0\Xi}(t)+C_+^{\Xi}C_{n}d_{n}^{+\Xi}(t).\eqno(A.14c)$$

\noindent
Now the primary task is to calculate the values of $d_{m,n}^{iA}(t)$ for the different Hamiltonians of the three-level systems and hence to obtain the corresponding time-dependent amplitudes.
\pagebreak
\bibliographystyle{plain}

\pagebreak
\pagebreak
\begin{center}
{\bf \large FIGURE CAPTION}
\end{center}
\begin{flushleft}
{\bf Fig.1: The energy levels of the lambda, vee and cascade transition types with the energy levels arranged as $E_3>E_2>E_1$. For each configuration, three choices of the basis states are possible (shown for lambda system only.)}
\end{flushleft}
\begin{flushleft}
{\bf Fig.2: The time evolution of the atomic entropy and population inversion of the lambda system for Case-I (Red), Case-II (Green) and Case-III (Blue) with $g_1=.2, g_2=.1, \alpha_m=\sqrt{30}$ and $\alpha_n=\sqrt{20}$.}
\end{flushleft}
\begin{flushleft}
{\bf Fig.3: The time evolution of the atomic entropy and population inversion of the vee system for Case-I (Red), Case-II (Green) and Case-III (Blue) with the same values of $g_1, g_2, \alpha_m$ and $\alpha_n$ as for Fig.2.}
\end{flushleft}
\begin{flushleft}
{\bf Fig.4: The time evolution of the atomic entropy and population inversion of the equidistant cascade system for Case-I with $g=.1$ and $\alpha_n=\sqrt{35}$. The presence of two distinct type of collapse and revival are evident for $W^{\Xi}_{12}$ and $W^{\Xi}_{23}$ transitions only.}
\end{flushleft}
\begin{flushleft}
{\bf Fig.5: The time evolution of the atomic entropy and population inversion of the equidistant cascade system for Case-II for the same values of $g$ and $\alpha_n$ as for Fig.4. Here we have single collapse and a single revival time.}
\end{flushleft}
\begin{flushleft}
{\bf Fig.6: The time evolution of the atomic entropy and population inversion of the equidistant cascade system for Case-III with the same values of $g$ and $\alpha_n$ as for Fig.4. The plots are similar to those for Case-I.}
\end{flushleft}
\pagebreak
\begin{picture}(300,86)(0,0)

\Line(100,90)(200,90)

\Text(55,90)[]{$E_3^\Lambda=\hbar(\omega_1+\omega_2)$}

\ArrowLine(155,90)(170,40)

\Text(175,70)[]{$\Omega_2$}

\Line(100,40)(200,40)

\Text(64,40)[]{$E_2^\Lambda=-\hbar\omega_2$}

\ArrowLine(120,10)(150,90) \Text(125,50)[]{$\Omega_1$}

\Line(100,10)(200,10)

\Text(64,10)[]{$E_1^\Lambda=-\hbar\omega_1$}

\Text(247,110)[]{Choice-I}
\Text(250,90)[]{$|m-1,n,+>$}
\Text(257,40)[]{$|m-1,,n+1,0>$}
\Text(242,10)[]{$|m,n,->$}

\Text(340,110)[]{Choice-II}
\Text(342,90)[]{$|m,n-1,+>$}
\Text(330,40)[]{$|m,n,0>$}
\Text(350,10)[]{$|m+1,n-1,->$}

\Text(420,110)[]{Choice-III}
\Text(420,90)[]{$|n,m,+>$}
\Text(425,40)[]{$|m,n+1,0>$}
\Text(425,10)[]{$|m+1,n,->$}
\end{picture}
\vspace{.2cm}
\begin{center}
{\bf Fig.1a}
\end{center}
\vspace{4cm}

\begin{picture}(100,86)(0,0)
\Line(100,90)(200,90)
\Text(75,90)[]{$E_3^V=\hbar\omega_1$}
\ArrowLine(120,90)(140,10)
\Text(165,20)[]{$\Omega_2$}
\Line(100,40)(200,40)
\Text(75,40)[]{$E_2^V=\hbar\omega_2$}
\ArrowLine(145,10)(160,40)
\Text(120,60)[]{$\Omega_1$}
\Line(100,10)(200,10)
\Text(55,10)[]{$E_1^V=-(\omega_1+\omega_2)$}

\Line(230,90)(330,90)
\Text(355,90)[]{$E_3^\Xi=\hbar\omega_2$}
\ArrowLine(280,40)(280,90)
\Text(280,20)[]{$\Omega_1$}
\Line(230,40)(330,40)
\Text(370,40)[]{$E_2^\Xi=\hbar(\omega_1-\omega_2)$}
\ArrowLine(270,10)(270,40)
\Text(272,60)[]{$\Omega_2$}
\Line(230,10)(330,10)
\Text(362,10)[]{$E_1^\Xi=-\hbar\omega_1$}
\end{picture}

\vspace{.2cm}
\begin{center}
{\bf Fig.1b, c}
\end{center}


\begin{figure}[h]
\begin{center}
\rotatebox{0} {\includegraphics [width=16cm]{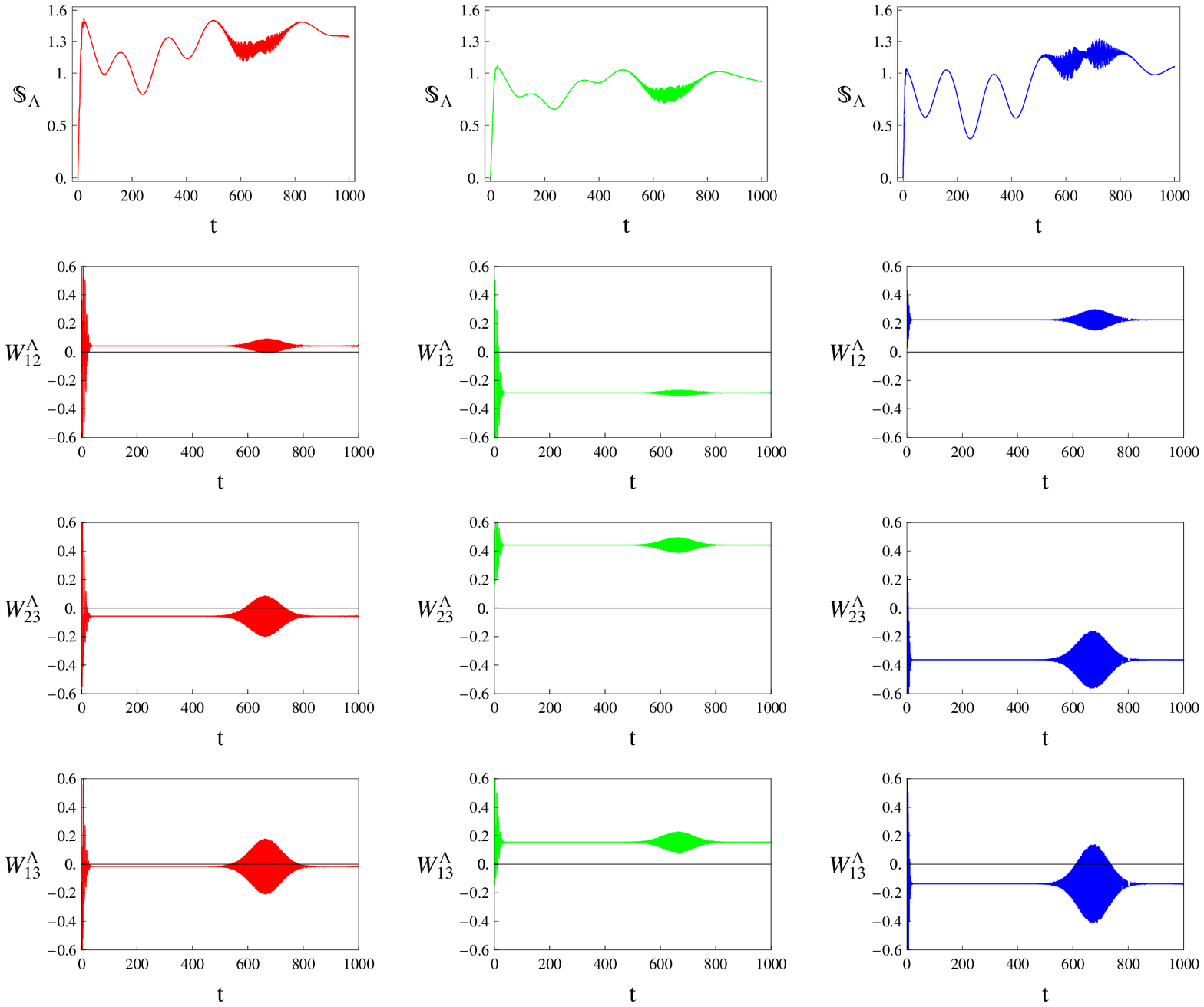}}
\end{center}
\centering
\noindent {\small {\bf Figure.2}}
\end{figure}

\begin{figure}[h]
\begin{center}
\rotatebox{0} {\includegraphics [width=16cm]{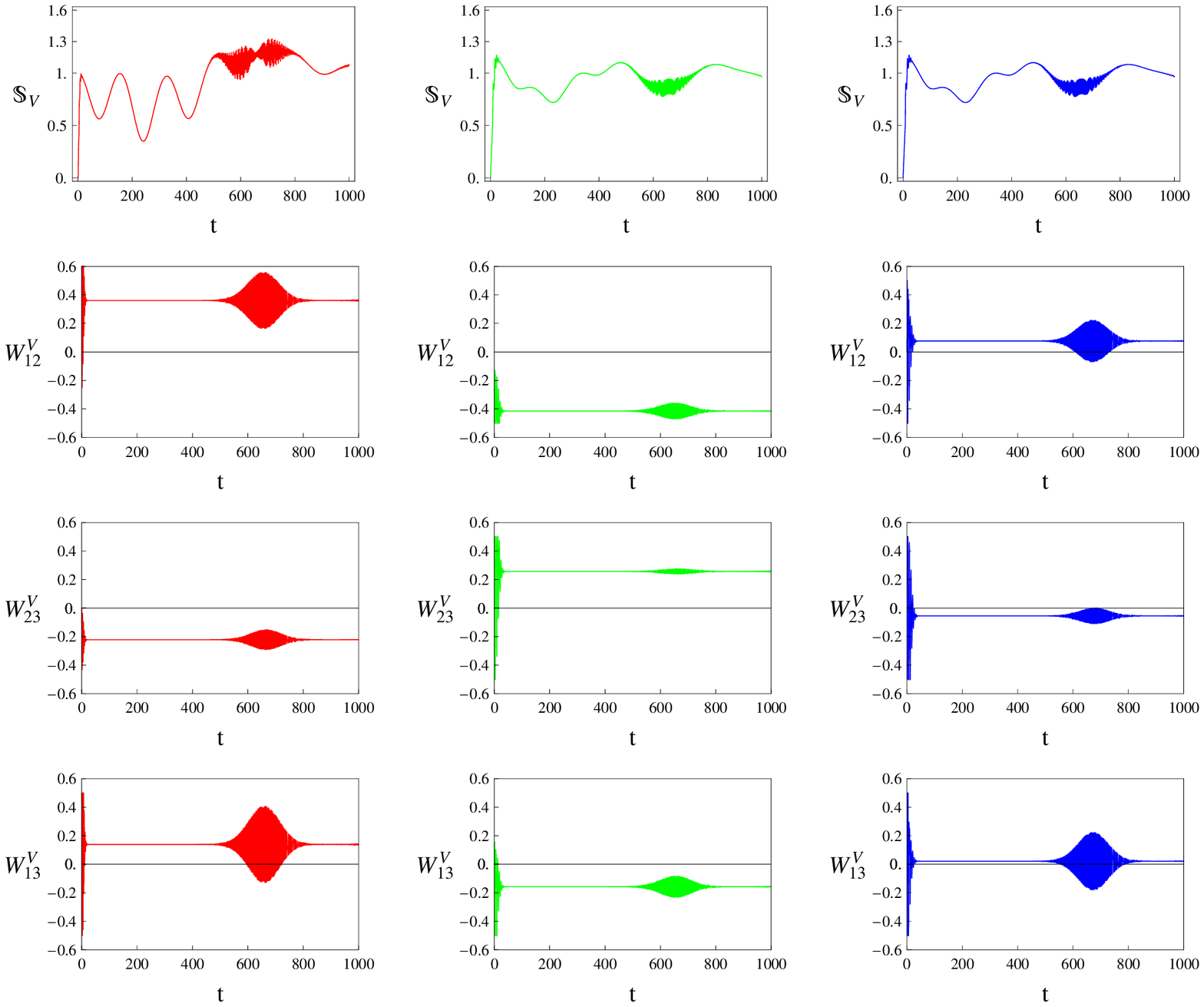}}
\end{center}
\centering
\noindent {\small {\bf Figure.2}}
\end{figure}

\begin{figure}[h]
\begin{center}
\rotatebox{0} {\includegraphics [width=8cm]{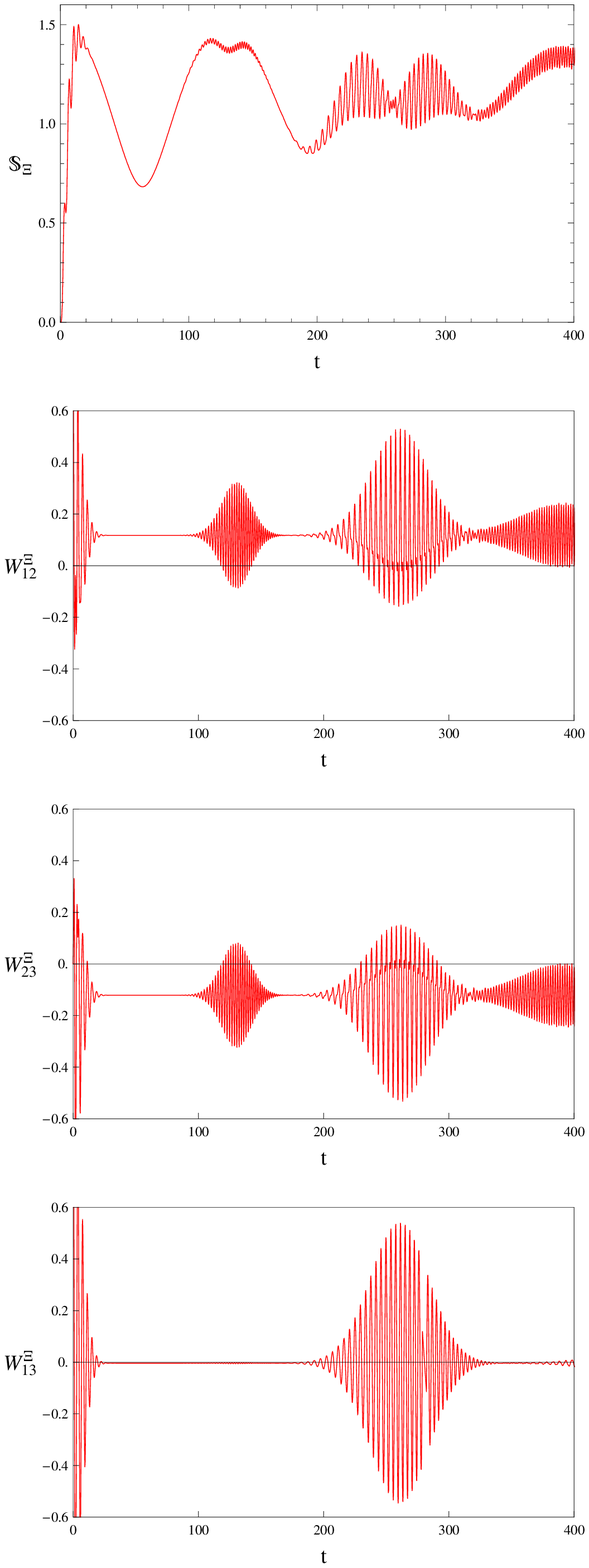}}
\end{center}
\centering
\noindent {\small {\bf Figure.4}}
\end{figure}

\begin{figure}[h]
\begin{center}
\rotatebox{0} {\includegraphics [width=8cm]{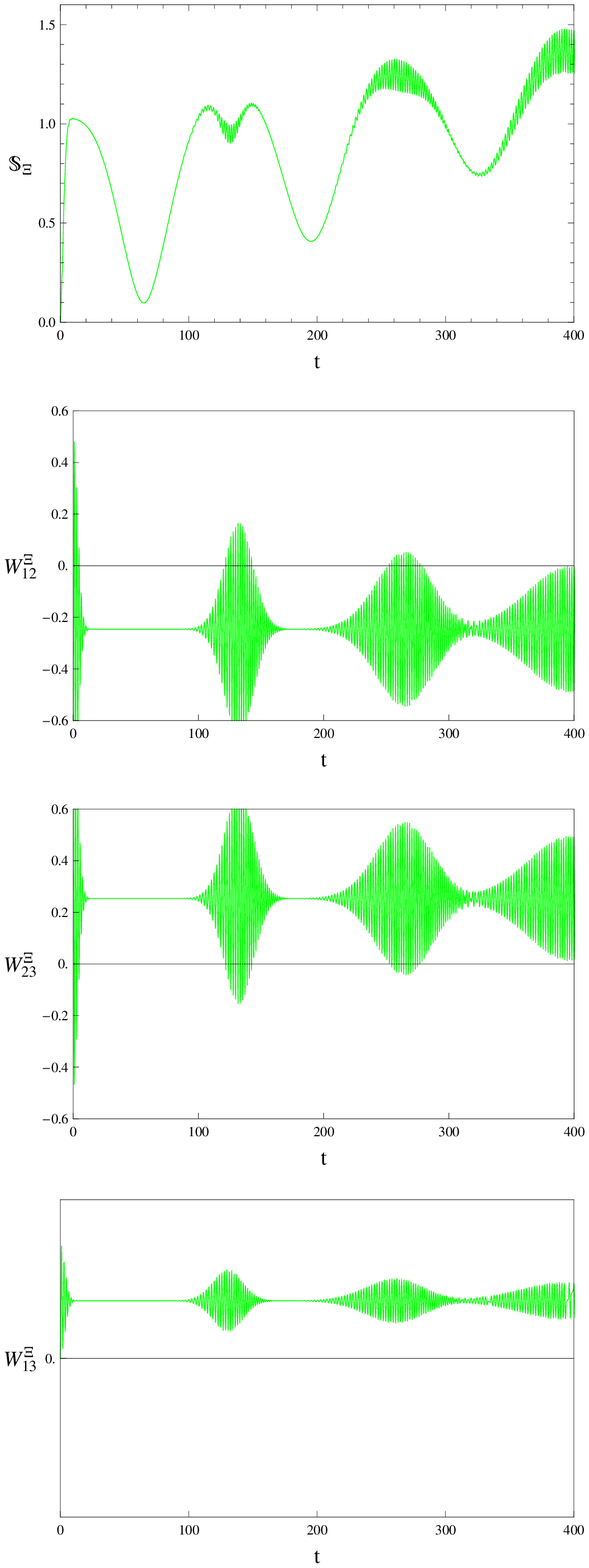}}
\end{center}
\centering
\noindent {\small {\bf Figure.5}}
\end{figure}

\begin{figure}[h]
\begin{center}
\rotatebox{0} {\includegraphics [width=8cm]{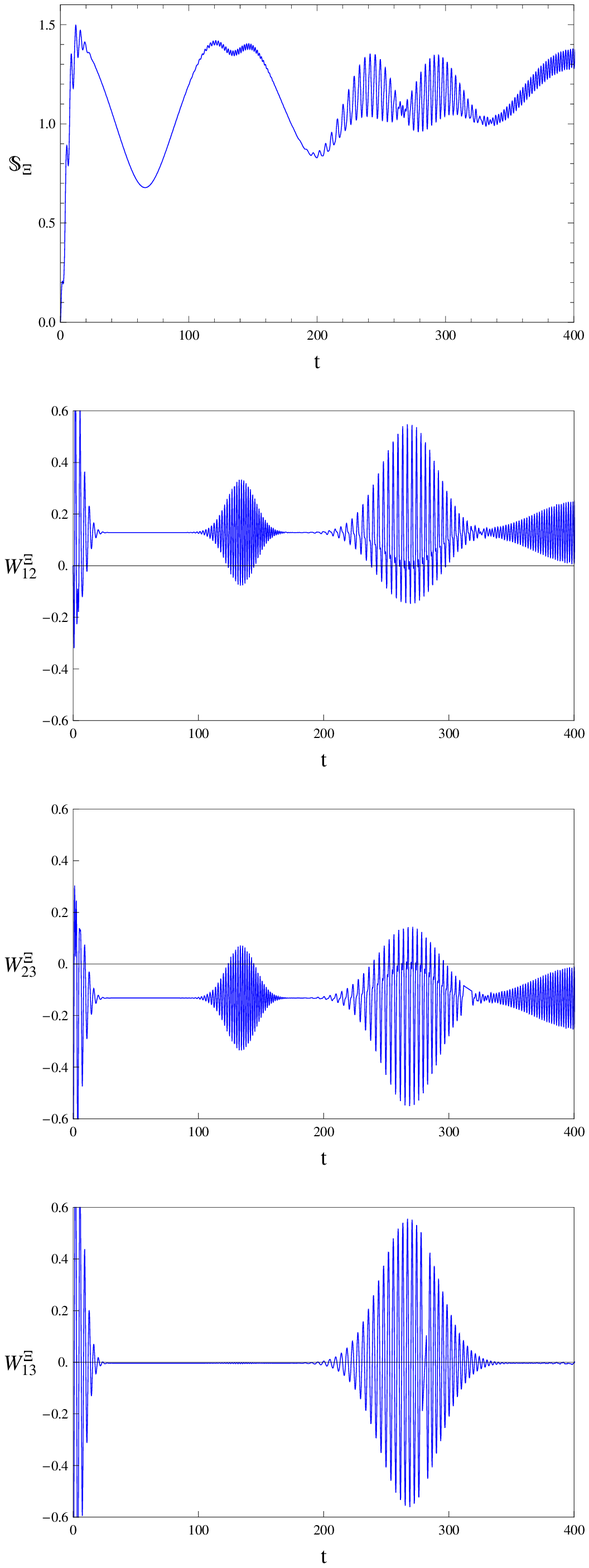}}
\end{center}
\centering
\noindent {\small {\bf Figure.6}}
\end{figure}

\end{document}